\documentclass[twocolumn,prd,tightenlines,nofootinbib,amsmath,amssymb,amsfonts,superscriptaddress,dvipsnames]{revtex4-1}
\usepackage{bm}
\usepackage{subeqnarray}
\usepackage{lipsum}
\usepackage{accents}
\usepackage{array}
\usepackage[normalem]{ulem}
\usepackage{tabularx}
\usepackage{enumerate}

\usepackage{pgfplots}
\usepackage{wasysym}
\usepackage{graphicx}
\usepackage{xcolor}
\usepackage{epsfig}
\usepackage[hidelinks,colorlinks=true,citecolor=Blue,linkcolor=Blue,urlcolor=Blue]{hyperref}

\newcommand{\be}{\begin{equation}}
\newcommand{\ee}{\end{equation}}
\newcommand{\bea}{\begin{eqnarray}}
\newcommand{\eea}{\end{eqnarray}}
\newcommand{\beas}{\begin{subeqnarray}}
\newcommand{\eeas}{\end{subeqnarray}}

\newcommand{\dd}{{\rm d}}

\newcommand{\com}[1]{}

\newcommand{\YP}{Y_{\rm P}}
\newcommand{\TFO}{T_{\rm FO}}
\newcommand{\TNuc}{T_{\rm Nuc}}
\newcommand{\He}[1]{{}^#1{\rm He}}
\newcommand{\Li}{{}^7{\rm Li}}
\newcommand{\Be}{{}^7{\rm Be}}

\newcommand{\abs}[1]{\left\lvert#1\right\rvert}

\DeclareMathOperator{\sign}{sgn}

\begin{document}

\title{Incomplete neutrino decoupling effect on big bang nucleosynthesis}

\author{Julien Froustey}
\email{froustey@iap.fr}
\affiliation{Institut d'Astrophysique de Paris, CNRS UMR 7095, Sorbonne
  Universit\'e, 98 bis Bd Arago, 75014 Paris, France}
\author{Cyril Pitrou}
\affiliation{Institut d'Astrophysique de Paris, CNRS UMR 7095, Sorbonne
  Universit\'e, 98 bis Bd Arago, 75014 Paris, France}
\date{\today}

\begin{abstract}
In the primordial Universe, neutrino decoupling occurs only slightly before electron-positron annihilations, leading to an increased neutrino energy density with order $10^{-2}$ spectral distortions compared to the standard instantaneous decoupling approximation. However, there are discrepancies in the literature on the impact it has on the subsequent primordial nucleosynthesis, in terms of both the magnitude of the abundance modifications and their sign. We review how neutrino decoupling indirectly affects the various stages of nucleosynthesis, namely, the freezing out of neutron abundance, the duration of neutron beta decay, and nucleosynthesis itself. This allows to predict the sign of the abundance variations that are expected when the physics of neutrino decoupling is taken into account. For simplicity, we ignore neutrino oscillations, but we conjecture from the detailed interplay of neutrino temperature shifts and distortions that their effect on final light element abundances should be subdominant. 
\end{abstract}

\maketitle


\section{Introduction}

The production of light elements during the first few minutes of our Universe, known as big bang nucleosynthesis (BBN), is a robust prediction of the standard cosmological model. The observational constraints on  $^4{\rm He}$~\cite{Izo14,Ave15} and deuterium abundances~\cite{Coo14,Coo16,Coo18} have now reached a percent-level precision, and the baryon abundance---which is the only free cosmological parameter which controls the synthesis---is also measured with percent precision from cosmic microwave background (CMB) anisotropies~\cite{Planck_2018}. In order to use BBN to constrain exotic cosmologies, or even to check the consistency of the theory with the one inferred from large-scale structure and CMB, it has thus become crucial to develop a theory of BBN that is much more precise than its associated observational constraints. Hence, we aim at least at a $10^{-3}$ precision level in the theory, and ideally even $10^{-4}$. The $^4{\rm He}$ abundance is essentially set by the neutron-to-proton ratio, which is in turn controlled by weak interaction rates. A comprehensive list of small physical effects, including radiative corrections, was developed in Refs.~\cite{Dicus1982,Lopez1997,LopezTurner1998,BrownSawyer,Serpico:2004gx} and reviewed in Ref.~\cite{Pitrou_2018PhysRept}, so as to reach a $0.1 \, \%$ theoretical precision on the weak rates. Numerical codes such as \texttt{PArthENoPE}~\cite{Parthenope,Parthenope_reloaded}, \texttt{AlterBBN}~\cite{alterbbn2012,alterbbn2018} and \texttt{PRIMAT}~\cite{Pitrou_2018PhysRept}, which were developed to predict these abundances, now incorporate these small physical effects, though with different approximations. The final abundances of other light elements, which are only at the level of traces, also depend directly on nuclear reaction rates, which themselves are also only known with a few percent precision in general, and can also be subject to radiative corrections~\cite{Pitrou:2019pqh}.

Among the small effects that affect these abundances is the incomplete decoupling of neutrinos prior to the reheating of photons
by electron-positron annihilations when the temperature of the Universe drops below $1 \, \mathrm{MeV}$. It leads to a small modification of the energy density in neutrinos~\cite{Dolgov1997,Esposito_NuPhB2000,Mangano2002,Mangano2005,Grohs2015,Escudero2019}, affecting the Hubble expansion rate. Therefore, a full treatment of the decoupling physics is required to properly describe the outcome of BBN. In this paper, we improve \texttt{PRIMAT}'s predictions by considering the detailed effects of incomplete neutrino decoupling. We choose to ignore the effect of neutrino oscillations, and focus instead on the effect of decoupling alone, as this will allow a physical understanding of how it influences final abundances. We comment further that from the understanding of the physics at play, it is expected that neutrino oscillations preserve the essential effects of neutrino decoupling, even though they alter the neutrino spectral distortions.

As far as we are aware, there have been three studies of the effect of decoupling on BBN abundances beyond the  $^4{\rm He}$ prediction, but they reached different conclusions as for the sign of abundance modifications. In Table 3 of Ref.~\cite{Mangano2005}, one can see that the $^4{\rm He}$ and $^7{\rm Li}$ abundances are increased, whereas the deuterium and $^3{\rm He}$ abundances are decreased, due to the incomplete decoupling of neutrinos. However, in Refs.~\cite{Grohs2015} and~\cite{Pitrou_2018PhysRept}, it was found that the variations are exactly in opposite directions for all abundances except $^4{\rm He}$. This is all the more surprising since Ref.~\cite{Pitrou_2018PhysRept} did not independently solve for the neutrino decoupling, but rather used the neutrino heating function of Ref.~\cite{Parthenope}. It should nevertheless be noted that the abundances of elements other than $\He4$ were not the main focus of Ref.~\cite{Mangano2005}, since precise measurements of the deuterium abundance were not available at that time.

The goal of this paper is to gain insight into the effects of the physics of neutrino decoupling on BBN, so as to understand in which sense, and to what extent, the abundances are affected. To that purpose, we have developed an independent implementation of the neutrino decoupling dynamical equations (without flavor oscillations), whose main ingredients and results for the neutrino spectra modifications are gathered in the next section, along with technical details in the Appendix. In Sec.~\ref{Sec:BBN} we review how final BBN abundances are modified by coupling these results to {\tt PRIMAT}.

Comparisons with respect to a fiducial cosmology, where neutrinos are artificially decoupled instantaneously prior to electron-positron annihilations, require the ability to map different homogeneous cosmologies. There is no unique way to perform this cosmology mapping, that is, to compute variations, exactly like how there is a gauge freedom when comparing a perturbed cosmology with a background cosmology. For instance, we can compare the fiducial instantaneous decoupling with the full neutrino decoupling physics, either using the same cosmological times or the same cosmological factors, or even the same plasma temperatures. The fact that there is no unique choice complicates the discussion of the physical effects at play, but the physical observables, e.g., the final BBN abundances, do not depend on it. We will systematically specify which variable is left constant (cosmic time, scale factor, or photon temperature) when comparing the true Universe to the fiducial one. Quantities written with a superscript $^{(0)}$ correspond to the fiducial (instantaneous decoupling) cosmology, and the variation of a quantity $\psi$ will be written as
\begin{equation}
\delta \psi \equiv \frac{\Delta \psi}{\psi^{(0)}} \equiv \frac{\psi - \psi^{(0)}}{\psi^{(0)}} \, .
\end{equation}

\section{Neutrino decoupling}
\label{Sec:Neutrinos}

\subsection{Neutrino kinetic equations}

The evolution of neutrino distribution functions is described by the Boltzmann equation
\begin{equation}
\label{eq:boltzmann_ini}
\left[\frac{\partial}{\partial t} - H p \frac{\partial}{\partial p}\right]f_{\nu_\alpha}(p,t) = C_{\nu_\alpha}[f_\nu,f_{e^\pm}] \, ,
\end{equation}
where $C_{\nu_\alpha}[f_j]$ is the collision term. This collision integral is dominated by two-body reactions, such as the annihilation process $\nu_\alpha + \bar{\nu}_\alpha \leftrightarrow e^- + e^+$ or neutrino-charged lepton scattering $\nu_\alpha + e^\pm \leftrightarrow e^\pm + \nu_\alpha$. The matrix elements for all relevant weak interaction processes are collected in, e.g., Ref.~\cite{Grohs2015} (see also Refs.~\cite{Hannestad_PhRvD1995,Dolgov1997}).

We consider the case without neutrino asymmetry, for which $f_{\nu_\alpha} = f_{\bar{\nu}_\alpha}$. We also assume that the distribution functions are the same for $\nu_\mu$ and $\nu_\tau$, since at the energy scales of interest (typically $\sim {\rm MeV}$), the muon and tau neutrinos have the same interactions with electrons and positrons. This is not true for electron neutrinos, which interact with the background $e^\pm$ via charged-current processes in addition to neutral-current channels. The set of equations is conveniently rewritten for numerical implementation in terms of comoving variables~\cite{Esposito_NuPhB2000,Mangano2005}:
\begin{itemize}
\item the normalized scale factor $x \equiv m_e/T_{\rm{cm}}$ (used in practice as an integration variable),
\item the comoving momentum $y \equiv p/T_{\rm{cm}}$, and
\item the dimensionless photon temperature $z \equiv T_\gamma/T_{\rm cm}$,
\end{itemize}
where the comoving temperature  $T_{\rm cm} \propto a^{-1}$ is only a convenient proxy for the scale factor \cite{Grohs2015}, and does not necessarily correspond to a physical temperature except at high values $T_{\rm cm} \gg 1 \ \rm{MeV}$ where all species are strongly coupled, and hence $T_\nu = T_\gamma = T_{\rm cm}$.

In the instantaneous decoupling approximation, neutrinos have an equilibrium Fermi-Dirac (FD) distribution at temperature $T_{\rm cm}$, which reads \begin{equation}
\label{eq:fnu_eq}
f_\nu^{(0)}(y) \equiv \frac{1}{e^y +1} \, .
\end{equation}
The charged leptons (electrons and positrons) are, in the range of temperatures of interest, kept in equilibrium with the plasma by fast electromagnetic interactions~\cite{Grohs2019}. Therefore, they follow a Fermi-Dirac distribution\footnote{The electron/positron dimensionless chemical potential $\mu_e/T_\gamma$ can be safely neglected as it is of the order of the baryon-to-photon ratio during most of the time of interest, which is smaller than $10^{-9}$; see, e.g., Fig. 30 in Ref.~\cite{Pitrou_2018PhysRept}.} at the plasma temperature $T_\gamma$, written as
\begin{equation}
f_{e^\pm} = \frac{1}{e^{\sqrt{p^2 + m_e^2}/{T_\gamma}}+1} = \frac{1}{e^{\sqrt{y^2 + x^2}/z}+1} \, .
\end{equation}

We need to solve for the evolution of the neutrino distribution functions and the photon temperature, i.e., the three variables $f_{\nu_e}(x,y)$, $f_{\nu_\mu}(x,y)$, and $z(x)$. In addition to the Boltzmann equations for neutrinos \eqref{eq:boltzmann_ini}, rewritten in the form
\begin{equation}
\label{eq:boltzmann_after}
\frac{\partial f_{\nu_\alpha}(x,y)}{\partial x} = \frac{1}{xH}C_{\nu_\alpha}(x,y) \, ,
\end{equation}
the third dynamical equation is the homogeneous energy conservation equation $\dot{\rho} = - 3 H (\rho + P) $. Following Ref.~\cite{Esposito_NuPhB2000}, it proves convenient for the stability of numerical implementations to introduce the dimensionless thermodynamic quantities
\begin{equation}
\bar{\rho} \equiv \rho \left(\frac{x}{m_e}\right)^4 \,, \qquad \bar{P} \equiv P \left(\frac{x}{m_e}\right)^4 \ .
\end{equation}
The energy conservation equation is then recast as an equation for $z(x)$ \cite{Esposito_NuPhB2000,Mangano2002}.

A comprehensive treatment of neutrino decoupling also requires taking into account two other effects. First, the electromagnetic interactions in the thermal bath of electrons, positrons, and photons lead to corrections with respect to vacuum quantum field theory. The plasma thermodynamics are modified through a change of the dispersion relations of $e^\pm$ and photons \cite{Heckler_PhRvD1994,Fornengo_PhRvD1997,LopezTurner1998}, which up to order $e^2$ can be described as a mass shift~\cite{Bennett2019}. These QED corrections to the energy density and pressure lead to corrective terms in the equation for $z(x)$, whose expressions were given in Ref.~\cite{Mangano2002}. In addition, they claimed that the modified dispersion relations must be introduced in the $f_{e^\pm}$ distribution functions, thus modifying the rates. However, and as pointed out in Ref.~\cite{Pitrou_2018PhysRept} for neutron/proton weak reactions, the mass shift is just part of the full finite-temperature radiative corrections for the weak rates derived in Ref.~\cite{BrownSawyer}. A comprehensive study of the finite-temperature corrections to neutrino weak rates is thus still needed, and we only include QED corrections in the plasma thermodynamics.\footnote{For completeness, we checked what happens if we include the mass shifts in the distribution functions, following Ref.~\cite{Mangano2002}. The results are identical at the level of precision considered, which is not surprising since computing collision integrals without the mass shift is already a first-order correction compared to the instantaneous decoupling case, where collision integrals vanish by definition.}

The second effect that needs to be taken into account is neutrino flavor oscillations, and it requires trading distribution functions for a density matrix formalism~\cite{SiglRaffelt,Stirner2018,Volpe_2013,Volpe_2015,Vlasenko_PhRevD2014,BlaschkeCirigliano}. The computation of collision integrals is considerably more demanding and has been performed over the last decade \cite{Mangano2005,Relic2016_revisited}. Nevertheless, understanding the physical phenomena involved in the effect of incomplete neutrino decoupling on primordial nucleosynthesis, even without oscillations, will serve as a guideline to predict the effect of oscillations based on the variation of the small number of quantities introduced in the following.

Throughout this paper, we will never consider neutrino oscillations, and QED corrections will not be included unless specified.

\subsection{Numerical implementation}

Two options have been considered to solve the kinetic equations. Either we use a discretization in momentum \cite{Dolgov1997,Grohs2015}, which is the only method with a reasonable computation time, or we expand the distribution functions in a basis of polynomials \cite{Esposito_NuPhB2000} so as to avoid the extrapolation of distribution functions between binning points. In order to combine both advantages we developed a ``hybrid" method: a binning in momentum is used within the collision integrals, which can therefore be efficiently computed using Simpson's method. Yet, rather than storing the values of $f_{\nu}$ for each discrete $y$ and interpolating between those points when needed, we perform an expansion over orthonormal polynomials. Namely, the neutrino distribution function is separated into an FD equilibrium one and a distortion according to
\begin{equation}
\label{eq:param_distortions}
f_{\nu_\alpha}(x,y) = \frac{1}{e^y + 1}\left[1 + \delta f_{\nu_\alpha}(x,y) \right] \, .
\end{equation}
We then expand $\delta f_{\nu_\alpha}$ in a set of polynomials,
\begin{equation}
\label{eq:expand_poly}
\delta f_{\nu_\alpha}(x,y) = \sum_{i=0}^{\infty}{a_i^{\alpha}(x) P_i(y)} \simeq \sum_{i=0}^{3}{a_i^{\alpha}(x) P_i(y)} \, ,
\end{equation}
where the polynomials $P_i$ are orthonormal with respect to the FD weight,
\begin{equation}
\label{eq:ortho}
\int_{0}^{\infty}{\dd y \frac{1}{e^y + 1} P_i(y) P_j(y)} = \delta_{ij} \, .
\end{equation}
The numerical results indicate (in agreement with Refs.~\cite{Esposito_NuPhB2000,Mangano2002}) that going up to order-3 polynomials is sufficient for our level of precision. Using the expansion \eqref{eq:expand_poly}, the Boltzmann equation \eqref{eq:boltzmann_after} becomes
\begin{equation}
\label{eq:da_dx}
\frac{d a_i^{\alpha}(x)}{dx} = \frac{1}{xH}\int_{0}^{\infty}{\dd y_1 \, P_i(y_1) \, C_{\nu_\alpha}(x,y_1)} \, .
\end{equation}
The explicit expression for $C_{\nu_\alpha}$ and the differential equation for $z(x)$ are collected in the Appendix~\ref{App:equations}.

The initial time of integration is a compromise, as it must be early enough to capture all of the relevant features of decoupling, but late enough such that weak rates are not too large, which would result in numerical stiffness when evaluating the collision integrals. We follow Refs.~\cite{Dolgov1997,Esposito_NuPhB2000} and take an initial (comoving) temperature $T_\mathrm{cm}^\mathrm{(in)} = 10 \ \mathrm{MeV}$, which corresponds to $x_\mathrm{in} = 0.0511$.
	
Neutrinos are kept in thermal equilibrium with the electromagnetic plasma before $x_\mathrm{in}$, so initially they have a FD distribution at the photon temperature, that is,
\begin{equation}
\label{eq:fnu_in}
f_{\nu}^{\mathrm{(in)}}(y) = \frac{1}{e^{p/T_\gamma^{\mathrm{(in)}}}+1} = \frac{1}{e^{y/z_\mathrm{in}}+1} \, ,
\end{equation}
which determines the initial values of the coefficients $a_i^\alpha$. Note that since electrons and positrons are not fully relativistic at $T_\mathrm{cm}^\mathrm{(in)}$, $z_{\rm in}$ is not exactly 1. By writing the entropy conservation of the full system of electrons, positrons, neutrinos, antineutrinos, and photons, one infers that $z_\mathrm{in} = 1.00003$, as in Ref.~\cite{Dolgov_NuPhB1999}. Besides, we checked that QED corrections to the plasma thermodynamics do not change $z_\mathrm{in}$ at this level of precision.

In agreement with previous statements in the literature \cite{Dolgov1997,Mangano2005,Grohs2015}, we find that a binning in momentum $y$ with at least 100 points in the range $[0,20]$ is sufficient to ensure convergence. Specifically, we chose a grid of 150 equally spaced points between $y_\mathrm{min} = 0.1$ and $y_\mathrm{max} = 20$. The integration variable $x$ ranges from $x_\mathrm{in} = 0.0511$ to $x_\mathrm{fin} \simeq 60$, where decoupling is essentially over.

\subsection{Results for neutrino transport}
\label{subsec:neutrino_results}

In this section we discuss the results obtained concerning neutrino decoupling. After reviewing some standard features (such as the plasma temperature and neutrino spectra), we introduce a parametrization that will be useful for studying the consequences on big bang nucleosynthesis.

\vspace{1cm}

\subsubsection{Overview}

The final dimensionless photon temperature is $z_\mathrm{fin} \simeq 1.3991$ (without QED corrections), which must be contrasted with the instantaneous decoupling value $z_0 = (11/4)^{1/3} \simeq 1.40102$. As expected, $e^+ e^-$ annihilations partly heat the neutrinos and the electromagnetic plasma is consequently less reheated. Including QED corrections, we get $z_\mathrm{fin} \simeq 1.3979$. These values are in very good agreement with previous results\footnote{Grohs {\it et al.}~\cite{Grohs2015} obtained a lower value for $z_\mathrm{fin}$ when including QED corrections, but this is due to an incorrect ``nonperturbative" implementation, as pointed out in Refs.~\cite{Pitrou_2018PhysRept,Bennett2019}. It was corrected in Ref.~\cite{Grohs_insights}.} \cite{Grohs2015,Relic2016_revisited}.

\begin{figure}[ht]
	\centering
	\includegraphics[trim=0cm 0.4cm 0.1cm 0.3cm, clip, width=8.6cm]{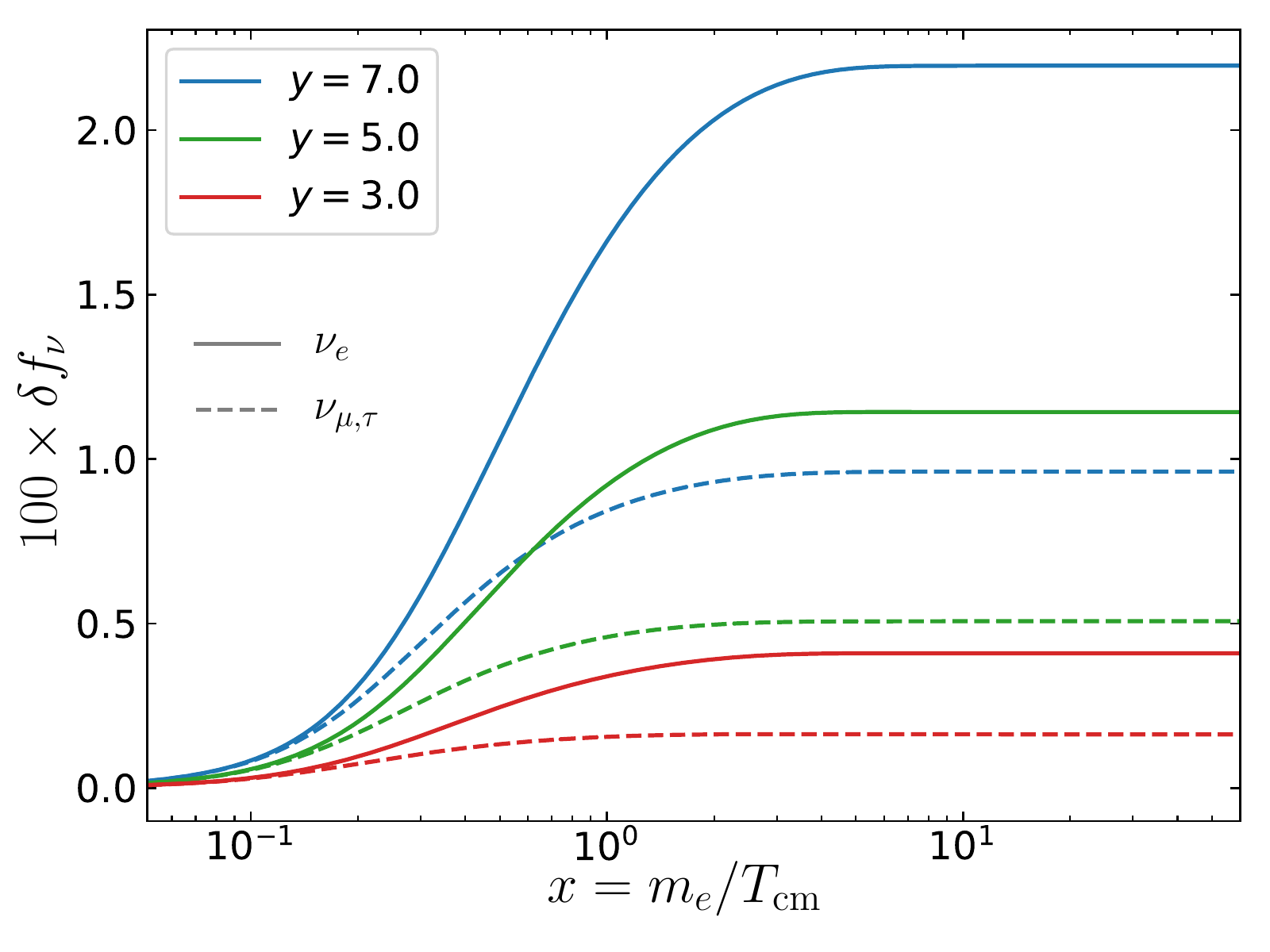}
	\caption{\label{fig:dfnu} Evolution of the distortion $\delta f_\nu$ as a function of $x$. From bottom to top: $y=3$, $5$, and $7$. Solid (dashed) lines correspond to electron (muon/tau) neutrinos. It agrees with Fig.~1 in Ref.~\cite{Grohs2015}, Fig.~3 in Ref.~\cite{Dolgov1997}, and Fig.~4 in Ref.~\cite{Esposito_NuPhB2000}.}
\end{figure}

The distortions with respect to the equilibrium Fermi-Dirac distribution are displayed in Fig.~\ref{fig:dfnu}, where we plot $\delta f_\nu$ [as defined in Eq.~\eqref{eq:param_distortions}] for comoving momenta $y=3$, $5$, and $7$. Due to the contribution of charged-current processes, $\nu_e$ distortions are enhanced with respect to those for $\nu_\mu$  and  $\nu_\tau$, and the  associated freeze-out occurs later. Note that neutrino distortions are of order $\delta f_\nu \sim 10^{-2}$, which is considerably larger than CMB spectral distortions~\cite{Lucca:2019rxf}.

Accordingly, we observe an increase of the energy density of neutrinos,
\begin{equation}
\delta \rho_{\nu_\alpha} \equiv \frac{\rho_{\nu_\alpha} - \rho_{\nu_\alpha}^{(0)}}{\rho_{\nu_\alpha}^{(0)}} \, ,
\end{equation}
with asymptotic values $\delta \rho_{\nu_e} \simeq 0.93 \, \%$ and $\delta \rho_{\nu_{\mu,\tau}} \simeq 0.39 \, \%$ (since these are frozen-out values, it is equivalent to computing them at constant $x$ or $T_\gamma$), which are still in excellent agreement with previous results. Through the Friedmann equation the expansion rate of the Universe is consequently modified, which has important consequences for primordial nucleosynthesis.

\subsubsection{Effective description of neutrinos}

\begin{figure}[!ht]
	\centering
	\includegraphics[trim=0cm 0.4cm 0.1cm 0.3cm, clip, width=8.6cm]{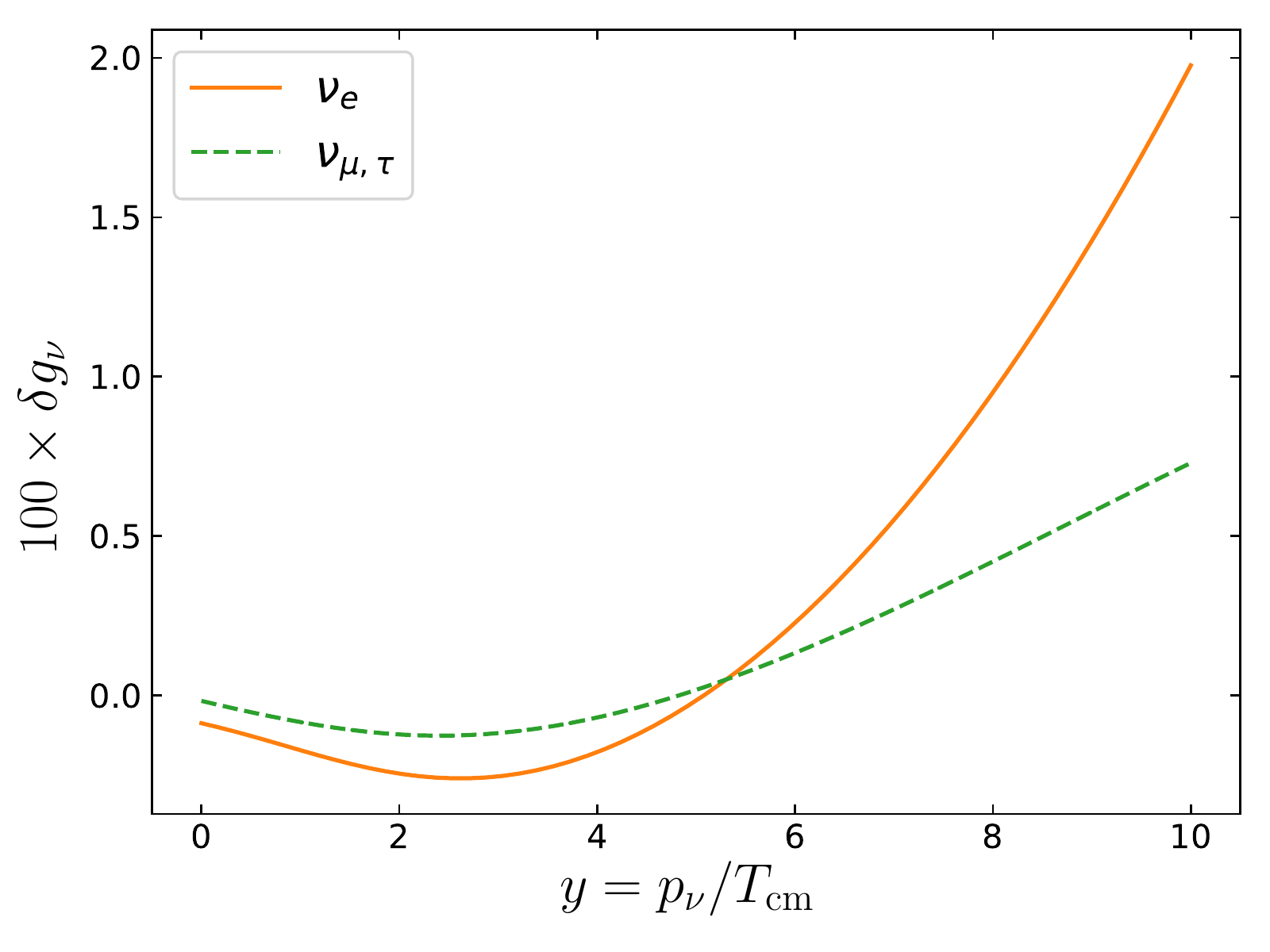}
	\caption{\label{fig:final_dist} Distortions with respect to the effective FD spectrum, as defined in Eq.~\eqref{eq:def_deltafhat}. Solid line: electron neutrinos. Dashed line: muon/tau neutrinos.}
\end{figure}

To scrutinize the precise role of neutrinos in BBN, it is particularly important to use a parametrization that separates the different effects of incomplete decoupling. To this end, we define an effective neutrino temperature $T_\nu$ (there is no genuine temperature since the distribution is not at equilibrium) as the temperature of the FD distribution with zero chemical potential which would have the same energy density as the real distribution, that is,
\begin{equation}
\label{eq:eff_z}
\rho_{\nu_\alpha} \equiv \frac78 \frac{\pi^2}{30} T_{\nu_\alpha}^4 \ \iff \ \bar{\rho}_{\nu_\alpha} \equiv \frac78 \frac{\pi^2}{30} z_{\nu_\alpha}^4 \, .
\end{equation}
Distortions are then defined with respect to this FD spectrum, according to
\begin{equation}
f_{\nu_\alpha}(x,y) = \frac{1}{e^{y/z_{\nu_\alpha}(x)}+1}\left[1 + \delta g_{\nu_\alpha}(x,y)\right] \, . \label{eq:def_deltafhat}
\end{equation}
By definition, these effective distortions are constrained so that Eq.~\eqref{eq:eff_z} holds; hence,\footnote{This approach for defining distortions is different from the CMB spectral distortions which are computed numerically using a number density effective temperature, rather than an energy density effective temperature~\cite{Lucca:2019rxf}. In the neutrino case, and given the size of distortions (which are much larger than for CMB), the neutrino energy density is more convenient since it enters directly into the Friedmann equation governing the expansion rate.}
\begin{equation}
\label{eq:constraint_deltag}
\int_{0}^{\infty}{\dd y \, y^3 \frac{\delta g_{\nu_\alpha}(y)}{e^{y/z_{\nu_\alpha}}+1}} = 0\,.
\end{equation}
We plot the final effective distortions $\delta g_{\nu_\alpha}(x_{\rm fin} \simeq 60,y)$ as a function of momentum in Fig.~\ref{fig:final_dist}. Even though these distortions and $\delta f_{\nu_\alpha}$ (shown in Fig.~2 of Ref.~\cite{Grohs2015}) are defined with respect to different references, their overall shapes are similar. This is expected since $z_{\nu_{\alpha}} \simeq 1$, and compared to a purely thermal distribution there is a deficit of low-energy neutrinos because of interactions with the hotter electrons and positrons (hence the negative values of $\delta g_{\nu_\alpha}$ for $y \leq 5$).

The final values we obtain are $z_{\nu_e}^{\mathrm{fin}} \simeq 1.0023$ and $z_{\nu_{\mu,\tau}}^{\mathrm{fin}} \simeq 1.0010$, showing once more the higher reheating of electron neutrinos. The total neutrino energy density, taking into account neutrinos and antineutrinos, is
\begin{equation}
\bar{\rho}_\nu = 2 \times \frac78 \frac{\pi^2}{30} \times \left( z_{\nu_e}^4 + 2  z_{\nu_\mu}^4 \right) \equiv 3 \times \frac78 \frac{\pi^2}{15} \times \hat{z}_\nu^4 \, ,
\end{equation}
where we introduced the average effective temperature of neutrinos
\begin{equation}
\hat{z}_\nu \equiv  \left( z_{\nu_e}^4 + 2 \,  z_{\nu_\mu}^4 \right)^{\frac14} \iff \widehat{T}_\nu \equiv  \left( T_{\nu_e}^4 + 2 \,  T_{\nu_\mu}^4 \right)^{\frac14}  \, .
\end{equation}

Being based on the energy density, these effective temperatures are adapted to the computation of the Hubble expansion rate, since in the very early Universe it is determined by the total radiation energy density $\rho_\mathrm{rad} = \rho_\gamma + \rho_{e^\pm} + \rho_\nu$. In the instantaneous decoupling approximation, we simply have
\begin{equation}
\rho_\mathrm{rad}^{(0)}(T_{\rm cm}) = \left[1 + \frac78 \left(\frac{T_{\rm cm}}{T_\gamma^{(0)}}\right)^{4} 3\right] \rho_\gamma^{(0)} + \rho_{e^\pm}^{(0)} \, ,
\end{equation}
since $\rho_\gamma = (\pi^2/15)T_\gamma^4$. The departure from this standard picture has historically been parametrized through the {\it effective number of neutrino species} $N_{\rm eff}$, i.e., the number of instantaneously decoupled neutrino species that would give the same energy density:
\begin{equation}
\rho_\nu(T_{\rm cm}) =  \frac78 \left(\frac{T_{\rm cm}}{T_\gamma^{(0)}}\right)^{4} N_\mathrm{eff} \times \rho_\gamma \, ,
\end{equation}
where $T_\gamma^{(0)}(T_{\rm cm})$ is the photon temperature at a given scale factor in the instantaneous decoupling approximation. Note that we could also define these quantities as a function of $T_\gamma$:
\begin{equation}
\rho_\nu(T_{\gamma}) =  \frac78 \left(\frac{T_{\rm cm}^{(0)}}{T_\gamma}\right)^{4} N_\mathrm{eff} \times \rho_\gamma \, .
\label{eq:rhonuNeff}
\end{equation}
 Either way, $N_\mathrm{eff}$ can be expressed as
\begin{equation}
\label{eq:Neff}
N_\mathrm{eff} = 3  \left(\frac{\hat{z}_\nu z^{(0)}}{z}\right)^4 \, .
\end{equation}
The final values of all of these parameters are summarized in Table~\ref{TableNeutrinos}, with comparison to previous results.

\renewcommand{\arraystretch}{1.05}

\begin{table}[h]
	\centering
	\begin{tabular}{|p{3.8cm}|cccc|}
  	\hline 
Frozen values & $z$ & $z_{\nu_e}$&  $ z_{\nu_\mu}$ &  $N_{\rm eff}$ \\
  \hline \hline
   {\it No QED corrections} & & & & \\ 
  Instantaneous decoupling & $1.40102$ & $1.$ & $1.$ & $3.000$ \\
  Naples group \cite{Mangano2005} & $1.3990$ & $1.0024$  & $1.0011$ & $3.035$ \\
   Grohs {\it et al.}~\cite{Grohs2015} & $1.3990$ &  $1.0023$ & $1.0009$ &$3.034$ \\
  \textbf{This paper} & $\mathbf{1.3991}$ & $\mathbf{1.0023}$ & $\mathbf{1.0010}$ & $\mathbf{3.034}$  \\ \hline \hline
    {\it With QED corrections} & & & & \\ 
  Instantaneous decoupling & $1.39979$ & $1.$ & $1.$ & $3.011$ \\
  Naples group \cite{Relic2016_revisited} & $1.39784$ &  $1.0023$ & $1.0010$ & $3.045$  \\
   Grohs {\it et al.}~\cite{Grohs_insights} & $1.39782$ &   &  &$3.044$ \\
  This paper & $1.39791$ & $1.0023$ & $1.0010$ & $3.044$  \\ \hline 
\end{tabular}
	\caption{Comparison of neutrino transport results with previous studies. We have converted energy density increases in effective neutrino temperatures via $\delta \rho_{\nu_\alpha} = z_{\nu_\alpha}^4 - 1$. For QED corrections, we took the most recent values from the Naples group \cite{Relic2016_revisited} and Grohs {\it et al.}~\cite{Grohs_insights} (instead of Ref.~\cite{Grohs2015}, where they incorrectly implemented these corrections).
	\label{TableNeutrinos}}
\end{table}

\vspace{2cm}

\begin{figure}[!ht]
	\centering
	\includegraphics[trim=0.1cm 0.1cm 0.1cm 0cm, clip, width=8.6cm]{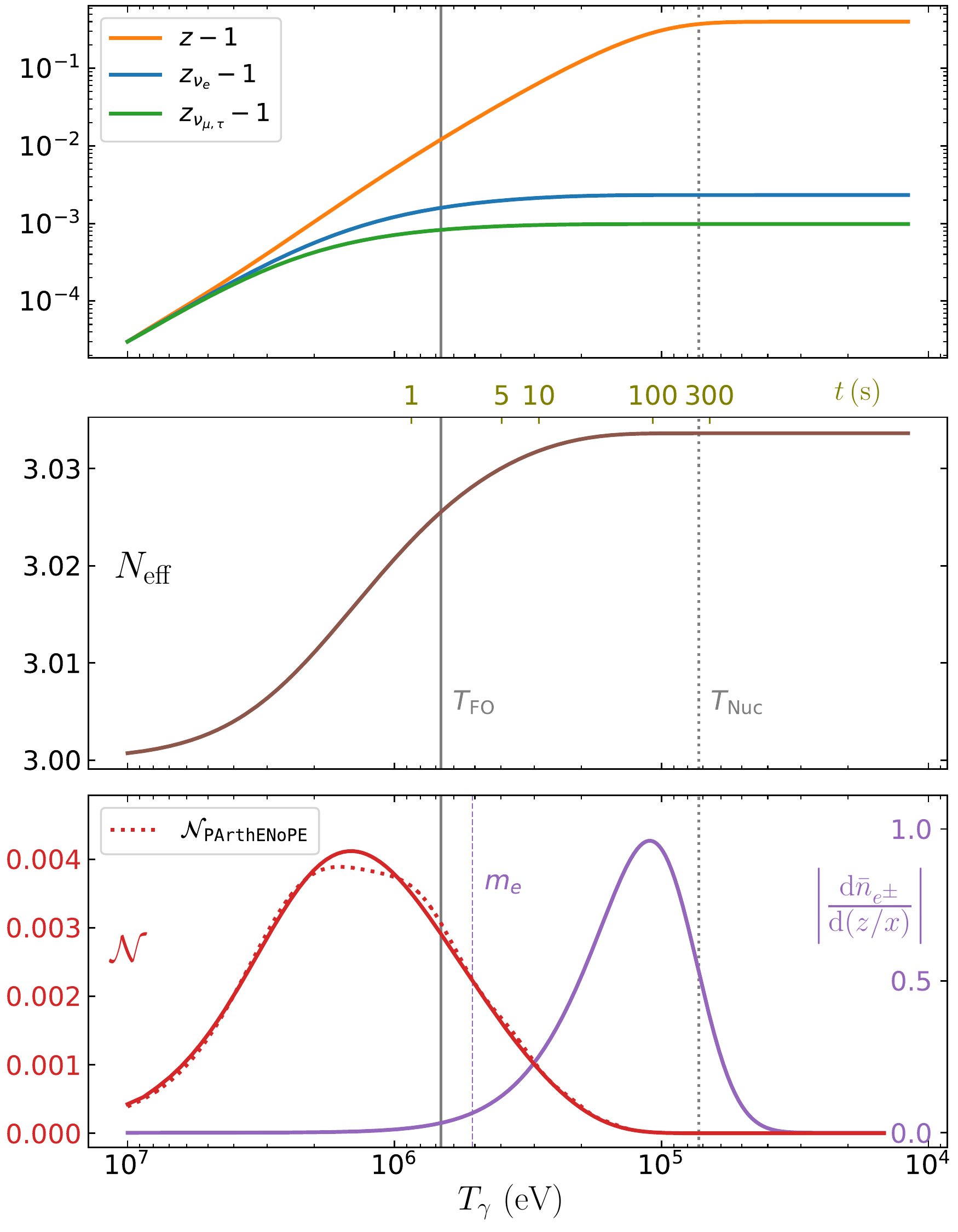}
	\caption{\label{fig:summary} Evolution of relevant quantities for neutrino decoupling, as a function of the plasma temperature. Top: Comoving (effective) temperatures of the plasma and neutrinos. Middle: Effective number of neutrinos, as defined in Eq.~\eqref{eq:Neff}. Bottom: Neutrino heating rate and variation of the comoving electron+positron density (derivative taken with respect to $z/x = T_\gamma/m_e$).}
\end{figure}

With this numerical simulation, we are able to grasp the variety of processes in place during the MeV age, summarized in Fig.~\ref{fig:summary}, where quantities are plotted with respect to the plasma temperature. The reheating of the different species is due to the entropy transfer from electrons and positrons, which is visualized by plotting the variation of their number density. For $T_\gamma \gg m_e$, electrons are relativistic and $\bar{n}_{e^\pm} \equiv (n_{e^-} + n_{e^+})\times (x/m_e)^3$ is constant, while for $T_\gamma \ll m_e$ the density drops to zero. The variation between those two constants corresponds to the annihilation period, which indeed starts around $T_\gamma \sim m_e$ and is over for $T_\gamma \sim 30 \, \mathrm{keV}$. At the beginning of this period, neutrinos progressively decouple and there is a heat transfer from the plasma, visualized through the dimensionless heating rate \cite{Parthenope,Parthenope_reloaded,Pitrou_2018PhysRept}
\begin{align}
\label{eq:Nheating}
\mathcal{N}(z) &= \frac{1}{z^4} \left(x \frac{d \bar{\rho}_\nu}{d x}\right)_{x=x(z)}  \nonumber \\
&= \frac{1}{z^4} \frac{1}{\pi^2 H} \int_0^{\infty}{\dd y \, y^3 \left[C_{\nu_e} + 2 C_{\nu_\mu}\right]} \, .
\end{align}
It is nonzero precisely during the decoupling of neutrinos. The slight overlap between the two curves in the bottom panel of Fig.~\ref{fig:summary} is the very reason why neutrinos are partly reheated. Finally, we plot the evolution of $N_\mathrm{eff}$, from $3$ before the MeV age to its frozen value $3.034$ (without QED corrections). Comparing with Fig.~5 in Ref.~\cite{Grohs2015}, we note that there is no ``plateau" before the freeze-out. This behavior can be considered as an artifact due to plotting $N_{\rm eff}$ as a function of $x=m_e/T_{\rm cm}$: the plateau is due to the difference between $T_{\rm cm}$ and $T_{\rm cm}^{(0)}$ for a given $T_\gamma$, and does not represent a meaningful physical effect (see also Fig.~7 in Ref.~\cite{Esposito_NuPhB2000}).

\section{Consequences for big bang nucleosynthesis}
\label{Sec:BBN}
By modifying the expansion rate of the Universe and affecting the neutron/proton weak reaction rates, incomplete neutrino decoupling will slightly modify the BBN abundances of light elements \cite{Pitrou_2018PhysRept,Mangano2005,Grohs2015}. We incorporate the results of Sec.~\ref{Sec:Neutrinos} into the BBN code \texttt{PRIMAT} to investigate the associated modification of abundances.

If $n_i$ is the volume density of isotope $i$ and $n_b$ is the baryon density, we define the \emph{number} fraction of isotope $i$, $Y_i \equiv n_i/n_b$. The \emph{mass} fraction is therefore $X_i \equiv A_i Y_i$, where $A_i$ is the nucleon number. It is customary to define $\YP \equiv X_{\He4}$ and $i/{\rm H} \equiv Y_i/Y_{\rm H}$.

To get a clear understanding of the physics at play, it is useful to recall the standard picture of BBN \cite{Peter_Uzan}.
\begin{enumerate}
\item Neutrons and protons track their equilibrium abundances, 
\begin{equation}
\label{eq:nse}
\left. \frac{n_n}{n_p}\right|_{\rm eq} = \exp{(-\Delta/T_\gamma)} \,,
\end{equation}
where $\Delta = m_n - m_p \simeq 1.293 \, \mathrm{MeV}$ is the difference of nucleon masses, until the so-called ``weak freeze-out," when the rates of $n \leftrightarrow p$ reactions drop below the expansion rate,
	\begin{equation}
	\label{eq:defofFO}
	\gamma \equiv \left. \frac{\Gamma_{n \to p} + \Gamma_{p \to n}}{H}\right|_{\TFO} \simeq 1\,.
	\end{equation}
\item After the freeze-out, neutrons only undergo beta decay until the beginning of nucleosynthesis, and a good approximation is
	\begin{equation}
	\label{eq:modelFO}
	X_n(\TNuc) = X_n(\TFO) \times \exp{\left[- \frac{t_{\rm Nuc} - t_{\rm FO}}{\tau_n}\right]}\, ,
	\end{equation}
	where $\tau_n \simeq 879.5 \, \mathrm{s}$ is the neutron mean lifetime. The nucleosynthesis temperature is usually defined when the {\it deuterium bottleneck} is overcome, with the criterion $n_D/n_b \sim 1$ \cite{Peter_Uzan,Neutrino_Cosmology}. It can also be associated with the maximum in the evolution of the deuterium abundance \cite{bernstein1989} which coincides with the drop in the density of neutrons (converted into heavier elements). We will adopt this definition, which is very close to the other criterion. Note that $t_{\rm Nuc} - t_{\rm FO} \simeq t_{\rm Nuc}$, since $t_{\rm FO} \ll t_{\rm Nuc}$.
\item Almost all free neutrons are then converted into $\He4$, leading to
	\begin{equation}
	\YP \simeq 2 X_n(\TNuc)\, .
	\end{equation}
\end{enumerate}
This indicates where incomplete neutrino decoupling will intervene. Weak rates, and thus the freeze-out temperature, are modified through the changes in the distribution functions (different temperatures and spectral distortions $\delta g_{\nu_e}$). But the changes in the energy density will also modify the relation $t(T_\gamma)$, leaving more or less time for neutron beta decay and light element production. This is the so-called {\it clock effect}, originally discussed in Refs.~\cite{Dodelson_Turner_PhRvD1992,Fields_PhRvD1993}. In summary, the neutron fraction at the onset of nucleosynthesis is modified as 
\begin{align}
\delta X_n^{[\rm Nuc]} = \frac{\Delta X_n(\TNuc)}{X_n^{(0)}(\TNuc)} 
&= \frac{\Delta X_n(\TFO)}{X_n^{(0)}(\TFO)} - \frac{\Delta t_{\rm Nuc}}{\tau_n}  \nonumber \\
&\equiv \delta X_n^{[\rm FO]} + \delta X_n^{[\Delta t]} \, , \label{eq:deltaxn}
\end{align}
with $\Delta t_{\rm Nuc} \equiv t_{\rm Nuc} - t_{\rm Nuc}^{(0)}$ (we neglected the variation of $t_{\rm FO}$). For freeze-out ($\delta X_n^{[\rm FO]}$), it is a variation at constant $\gamma = 1$, which we take as our definition of freeze-out. $\delta X_n^{[\rm Nuc]}$ is the neutron abundance variation between the onset of nucleosynthesis in the ``actual" Universe and the one in the reference universe. Given our definition of $\TNuc$, the constant quantity here is $\dd X_D/\dd t = 0$.

Note that this model of freeze-out is quite similar to the instantaneous decoupling approximation for neutrinos, i.e., we condense a gradual process into a snapshot. Actually, in the range $3 \gtrsim \gamma \gtrsim 0.2$, there is a smooth transition between nuclear statistical equilibrium [Eq.~\eqref{eq:nse}] and pure beta decay. For the sake of argument, we keep the criterion $\gamma \simeq 1$, and we will point out the limits of this model in the following discussions when necessary.

\renewcommand{\arraystretch}{1.4}

\begin{table*}[!htb]
	\centering
	\begin{tabular}{|l|cccccc|}
  	\hline 
 BBN framework & $\YP$  & $\delta \YP \,  (\%)$ & ${\rm  D}/{\rm H}  \times 10^5$ &$\delta \left({\rm  D}/{\rm H}\right) \,(\%)$&  ${\left(\He3 + \mathrm{T}\right)}/{\rm  H} \times 10^5$ &  ${\left(\Li + \Be\right)}/{\rm  H} \times 10^{10}$ \\
  \hline \hline
  Inst. decoupling, no QED & $0.24262$ &  $0$ & $2.423$ &$0$& $1.069$  & $5.635$ \\ \hline
  $\widehat{T}_\nu$ & $0.24274$ &  $0.050$ & $2.433$ &$0.38$& $1.070$  & $5.613$ \\
    $T_{\nu_e},$ no distortions & $0.24266$ & $0.015$ & $2.432$ & $0.36$ & $1.070$  & $5.612$ \\
  $T_{\nu_e},$ with distortions & $0.24276$ &  $0.056$ & $2.433$ &$0.39$& $1.070$  & $5.613$ \\ \hline \hline
   Inst. decoupling, with QED & $0.24262$ &  $0$ & $2.426$ &$0$& $1.069$  & $5.627$ \\ \hline
  $\widehat{T}_\nu$ & $0.24274$ &  $0.050$ & $2.435$ &$0.38$& $1.070$  & $5.606$ \\
    $T_{\nu_e},$ no distortions & $0.24265$ & $0.015$ & $2.435$&$0.36$ & $1.070$  & $5.604$ \\
  $T_{\nu_e},$ with distortions & $0.24275$ &  $0.056$ & $2.435$ &$0.38$& $1.070$  & $5.606$\\ \hline 
\end{tabular}
	\caption{Light element abundances, at the Born approximation level, for various implementations of neutrino-induced corrections. See Sec.~\ref{Subsec:other_corrections} for results with the full corrections derived in Ref.~\cite{Pitrou_2018PhysRept}. Since tritium and $\Be$ decay into $\He3$ and $\Li$, respectively, their abundances are usually summed.
	\label{Table:Corrections}}
\end{table*}

\renewcommand{\arraystretch}{1.2}

\subsection{Incomplete neutrino decoupling in \texttt{PRIMAT}}

In the version of \texttt{PRIMAT} used in Ref.~\cite{Pitrou_2018PhysRept}, the lack of effective temperatures and spectral distortion values across the nucleosynthesis era required an approximate strategy to include incomplete neutrino decoupling. It consisted in neglecting spectral distortions $\delta g_{\nu} = 0$ while computing an effective average temperature $\widehat{T}_\nu$ from the heating rate \eqref{eq:Nheating}. The values of $\mathcal{N}$ were obtained from a fit given in \texttt{PArthENoPE} \cite{Parthenope}  [Eqs.~(A23)--(A25)], computed by Pisanti {\it et al.} from the results of Refs.~\cite{Mangano2002,Mangano2005}.

This method correctly captures the changes in the expansion rate (since the energy density is well computed from $\widehat{T}_\nu$), but \emph{a priori} it handles the weak rates poorly: electron neutrinos are too cold ($T_{\nu_e} > \widehat{T}_\nu$), and their spectrum is not distorted. This should in principle have consequences for the neutron-to-proton ratio at freeze-out, and thus on the final abundances.

We modified \texttt{PRIMAT} to introduce the results from neutrino transport analysis. Since the useful variable in nucleosynthesis is the plasma temperature $T_\gamma$, all other quantities ($x$, $T_{\nu_\alpha}$, $a_i^{\alpha}$) are interpolated. Depending on the options chosen, one can then use the ``real" effective neutrino temperatures or the average temperature for comparison with the previous approach (keeping the true total energy density in each case). The distortions $\delta g_{\nu_e}$ are computed thanks to the coefficients $a_i^e$, and they correct the weak rates at the Born level. Following the notations of Ref.~\cite{Pitrou_2018PhysRept} [Eq.~(76) and subsequent equations], we add the corrections
\begin{subequations}
\label{eq:deltagamma}
\begin{align}
\Delta \Gamma_{n \to p} &=  K \int_{0}^{\infty}{p^2 \mathrm{d} p \left[\delta \chi_+(E) + \delta \chi_+(-E)\right]} \, ,  \label{subeq:np} \\
\Delta \Gamma_{p \to n} &= K \int_{0}^{\infty}{p^2 \mathrm{d} p \left[\delta \chi_-(E) + \delta \chi_-(-E)\right]} \, , \label{subeq:pn}
\end{align}
\end{subequations}
where $K =  \left(m_e^{5} \lambda_0 \tau_n\right)^{-1}$, $E = \sqrt{p^2 + m_e^2}$ is the electron energy, and
\begin{align}
\delta \chi_{\pm} (E) &= \left(E_\nu^\mp\right)^2 f_{e}(-E) \frac{\sign{(E_\nu^\mp)} \times \delta g_{\nu_e}(\abs{E_\nu^\mp})}{e^{\abs{E_\nu^\mp}/T_{\nu_e}}+1} \, , \\
E_\nu^\mp &= E \mp \Delta \, .
\end{align}
The $\sign$ function accounts for the fact that $f_{\nu_e}(\abs{E_\nu^\mp})$ appears as part of a Pauli blocking factor if $E_\nu^\mp < 0$, i.e., the neutrino is in a final state.

Our results are summarized in Table~\ref{Table:Corrections}. We consider three different implementations:
\begin{itemize}
\item[\it (i)] The earlier \texttt{PRIMAT} approach (no distortions and an average neutrino temperature), with slight differences compared to Ref.~\cite{Pitrou_2018PhysRept} since our results are used instead of \texttt{PArthENoPE}'s results. We call this approach ``$\widehat{T}_\nu$" in Tables~\ref{Table:Corrections} and \ref{Table:Full_Corrections} and Figs.~\ref{fig:analyzevariations} and \ref{fig:analyzevariationsfull}.
\item[\it (ii)] The weak rates including the real electron neutrino temperature, but still without spectral distortions. We call this approach ``$T_{\nu_e},$ no distortions."
\item[\it (iii)] Full results from neutrino evolution. We call this approach ``$T_{\nu_e},$ with distortions."
\end{itemize}
Note that these three scenarios take place in identical cosmologies, with the \emph{same} energy density; using the proper $\nu_e$ temperature and including distortions only affect the weak rates. This emphasizes the particular role of spectral distortions: the most striking---and somehow unexpected---feature is the proximity of the results in cases {\it (i)} and {\it (iii)}, which is investigated further in the next section.

The results from previous implementations of incomplete neutrino decoupling in BBN codes are shown in Table~\ref{Table:Previous}, and we check that our results are in close agreement with Grohs {\it et al.}~\cite{Grohs2015}, but with opposite signs of variation (except for ${}^4{\rm He}$) compared to the results of Mangano {\it et al.}~\cite{Mangano2005}. The extensive study in the next section sheds a new light on the different phenomena involved.

\begin{table*}[!ht]
	\centering
	\begin{tabular}{|l|cccc|}
  	\hline 
 Variation of abundances & $\delta \YP$ & $\delta \left({\rm  D}/{\rm H}\right) $&  $ \delta \left({\left(\He3 + \mathrm{T}\right)}/{\rm  H}\right) $ &  $\delta \left({\left(\Li + \Be\right)}/{\rm  H}\right) $ \\
  \hline \hline
  \emph{No QED corrections} & & & &   \\ 
  Naples group \cite{Mangano2005} & $6.06 \times 10^{-4}$ &  &  &  \\
  Grohs \emph{et al.}~\cite{Grohs2015} & $4.636 \times 10^{-4}$ &  $3.686 \times 10^{-3}$ & $1.209 \times 10^{-3}$ &$-3.916 \times 10^{-3}$ \\
  This paper & $5.636 \times 10^{-4}$ & $3.869 \times 10^{-3}$ & $1.268 \times 10^{-3}$ & $-3.867 \times 10^{-3}$  \\ \hline \hline
    \emph{QED corrections included} & & & & \\ 
  Naples group \cite{Mangano2005} & $6.96 \times 10^{-4}$ &  $-2.8 \times 10^{-3}$ & $-1.0 \times 10^{-3}$ & $3.77 \times 10^{-3}$  \\
  This paper & $5.604 \times 10^{-4}$ & $3.831 \times 10^{-3}$ & $1.256 \times 10^{-3}$ & $-3.828 \times 10^{-3}$  \\ \hline 
\end{tabular}
	\caption{Comparison with previous results. Note that baseline values are different in the cases that do or do not include QED corrections (see Table~\ref{Table:Corrections}). The values given by the Naples group in Ref.~\cite{Mangano2005} are absolute variations, and we need the baseline values to compute relative variations; as these were not given, we use our own baseline values.
	\label{Table:Previous}}
\end{table*}

\subsection{Detailed analysis}

We now review the physics that allows us to understand the numerical results of Table~\ref{Table:Corrections}. We first detail the physics affecting the helium abundance, which is directly related to the neutron fraction at the onset of nucleosynthesis, before turning to the production of other light elements, for which the clock effect dominates.

\subsubsection{Neutron/proton freeze-out}

Previous articles \cite{Dodelson_Turner_PhRvD1992,Fields_PhRvD1993,Mangano2005} studied the variation of $n \leftrightarrow p$ rates due to incomplete neutrino decoupling at constant scale factor, claiming that $H$ was left unchanged at a given $x$. This argument of constant total energy density, namely $\Delta \rho_\nu = - \Delta \rho_{\rm em}$, requires $T_\gamma \simeq T_\nu$ (cf. Appendix 3 in Ref.~\cite{Dodelson_Turner_PhRvD1992}). However, by looking at the top panel of Fig.~\ref{fig:summary} it appears that at freeze-out $T_\gamma$ and $T_{\nu_{\alpha}}$ differ by $\sim 1 \, \%$, which is the typical order of magnitude of variations we are interested in. Moreover, the analysis of Ref.~\cite{Dodelson_Turner_PhRvD1992} used thermal-equivalent distortions of neutrinos spectra (i.e., only effective temperatures, no $\delta g_{\nu}$) and the numerical relation $\Delta X_n^{[\rm FO]} \simeq -0.1 \Delta T_i/T_i$, which requires separating the temperature variations of the different species, which seems inconsistent with the constant energy density requirement. Their results are nonetheless in good agreement with numerical results; however, our findings seem to indicate that the proper way to implement thermal-equivalent distortions is with a unique, average neutrino temperature, thus slightly modifying the arguments in Refs.~\cite{Fields_PhRvD1993,Mangano2005}.

Due to the rich interplay of the processes involved, an analytical estimate of $\delta X_n^{[\rm FO]}$ is particularly challenging. Since our goal is to provide a satisfying physical picture of the role of neutrinos in BBN, and thus to check Eq.~\eqref{eq:deltaxn}, we perform a numerical evaluation. 

\begin{figure}[!ht]
	\centering
	\includegraphics[trim=0.1cm 0.4cm 0.2cm 0.2cm, clip, width=8.6cm]{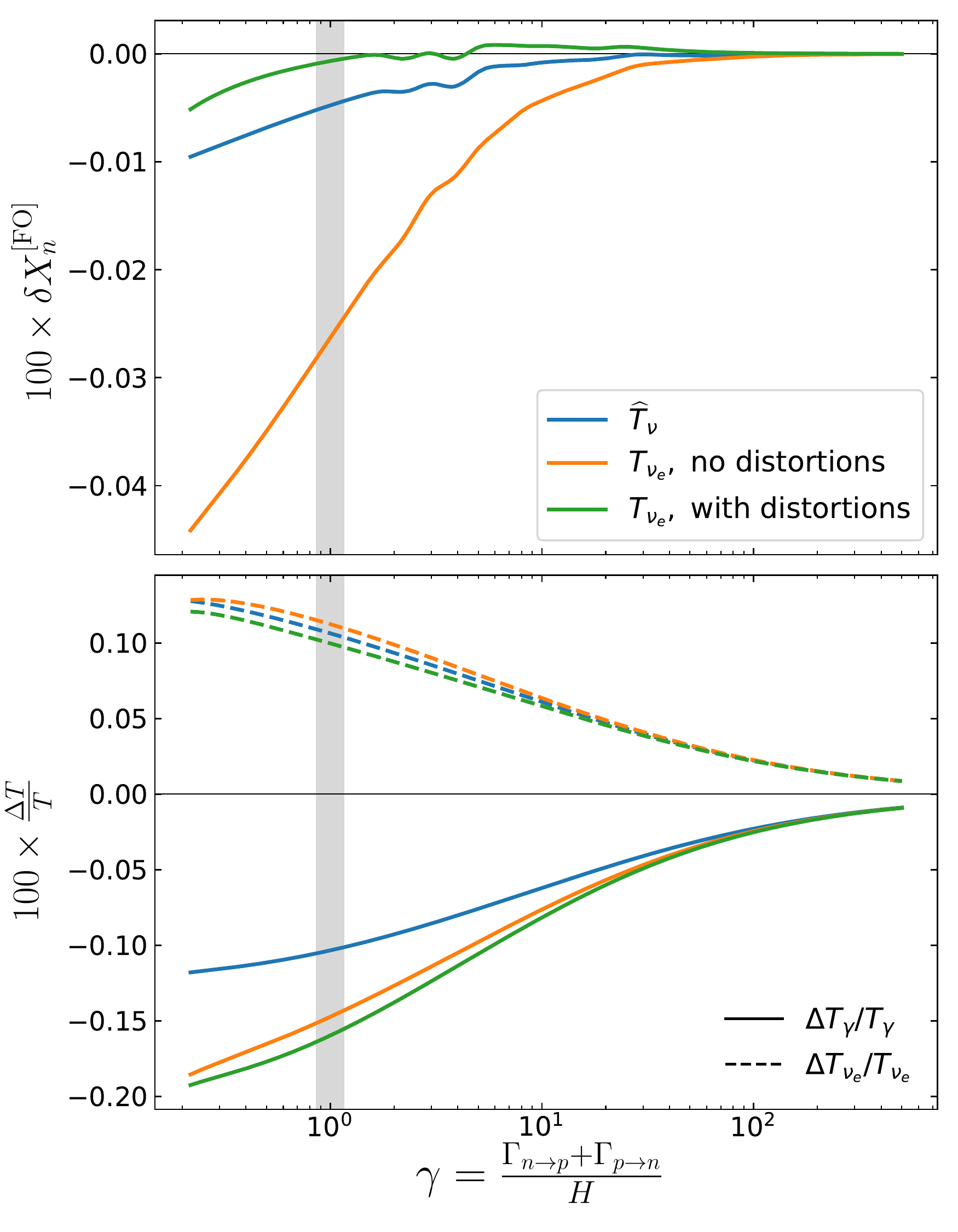}
	\caption{\label{fig:analyzevariations} Neutron fraction (top) and temperature (bottom) variations at freeze-out, in the different implementations of neutrino-induced corrections.}
\end{figure}

Figure~\ref{fig:analyzevariations} shows the variation of $X_n$ and $T_{\nu_e,\gamma}$ for the different implementations of neutrino-induced corrections around the time of freeze-out. In each case, incomplete neutrino decoupling leads to a decrease of $X_n$. We also find the interesting feature (already evidenced in Table~\ref{Table:Corrections}) that a thermal-equivalent approach (without distortions) with an average neutrino temperature gives results that are close to the full description.

For each implementation of neutrino-induced corrections the evolution of the photon temperature $z(x)$ is the same; the difference lies in whether or not we include $z_{\nu_e}$ and $\delta g_{\nu_e}$. But the quantities in Fig.~\ref{fig:analyzevariations} are plotted with respect to $\gamma$, which is a different function of $x$ in each case. For instance, when including the real $\nu_e$ temperature, weak rates increase and freeze-out is delayed, leading to a smaller $T_\gamma(\gamma \simeq 1) \equiv \TFO$: the orange curve is below the blue one in the bottom panel of Fig.~\ref{fig:analyzevariations}. Adding the distortions increases the rates even more, and slightly decreases $\TFO$ (green curve). One would then expect a reduction of $X_n$, which would track its equilibrium value longer. While this is true for thermal corrections (orange curve below the blue one in the top panel of Fig.~\ref{fig:analyzevariations}), adding the distortions disrupts this picture.

\begin{figure}[!ht]
	\centering
	\includegraphics[trim=0.1cm 0.4cm 0.2cm 0.2cm, clip, width=8.6cm]{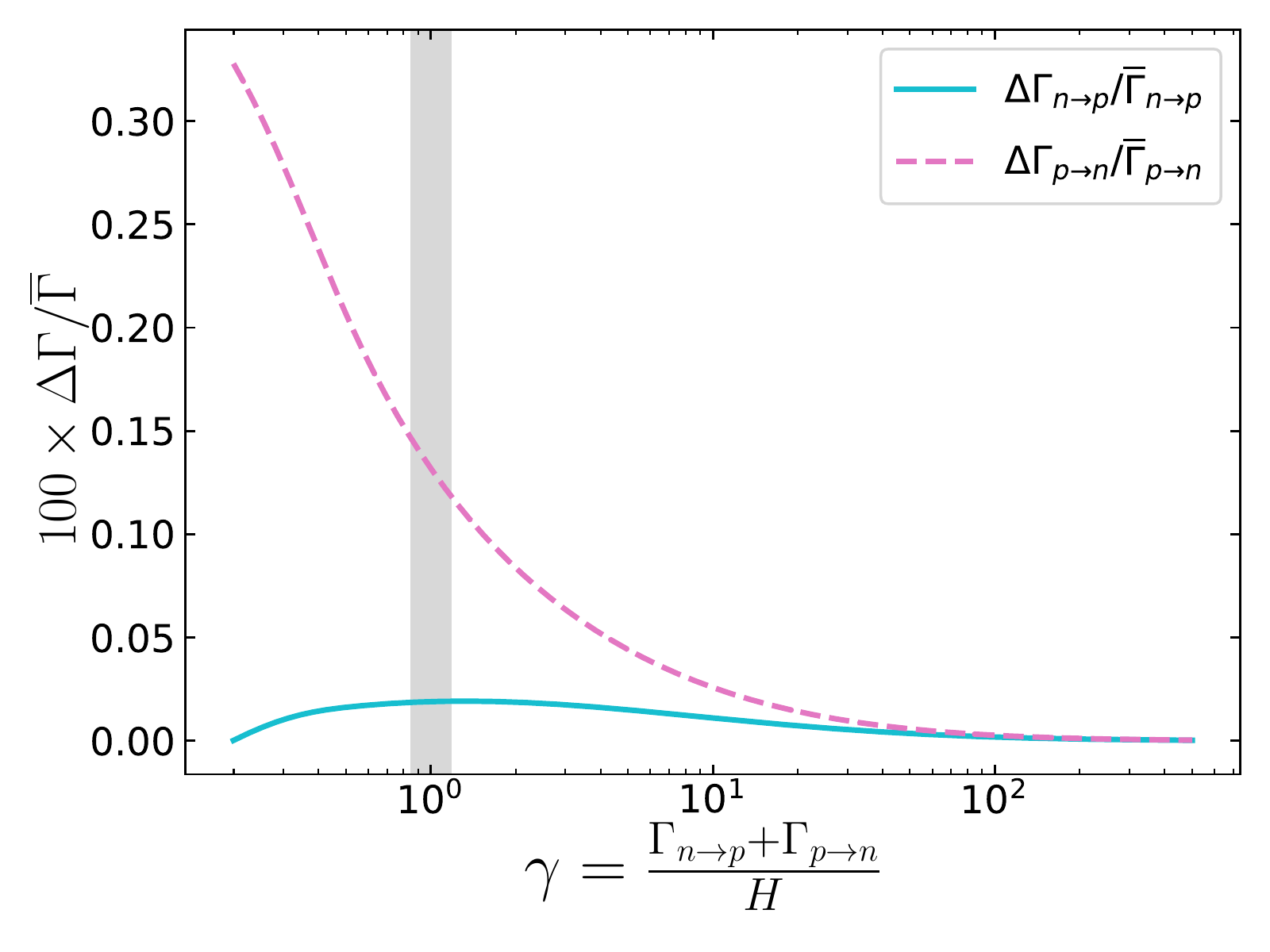}
	\caption{\label{fig:detailedbalance} Relative corrections to $n \leftrightarrow p$ weak rates, with $\Delta \Gamma_{n \leftrightarrow p}$ defined in Eq.~\eqref{eq:deltagamma}. To ensure detailed balance requirements, we enforce $T_{\nu_e} = T_\gamma$.}
\end{figure}

Indeed, the main effect of including neutrino spectral distortions is to alter the detailed balance relation $\overline{\Gamma}_{p \to n} = e^{-\Delta/T} \overline{\Gamma}_{n \to p}$. Let us parametrize this deviation from detailed balance as
\begin{equation}
\label{eq:detailedbalance}
\Gamma_{p \to n} = \exp{ \left(-\frac{\Delta}{T} + \sigma_\nu\right)} \Gamma_{n \to p} \, ,
\end{equation}
with $\sigma_\nu \ll 1$. Writing this in terms of the Born rates $\overline{\Gamma}$ (which satisfy the detailed balance equation), we get
\begin{equation}
\sigma_\nu = \frac{\Delta \Gamma_{p \to n}}{\overline{\Gamma}_{p \to n}} - \frac{\Delta \Gamma_{n \to p}}{\overline{\Gamma}_{n \to p}} \, ,
\end{equation}
leading to a change in the equilibrium neutron abundance,
\begin{equation}
\label{eq:dxnsigma}
\delta X_n^{(\mathrm{eq})} = (1 - X_n) \sigma_\nu \, ,
\end{equation}
since $X_n/(1-X_n) = n_n/n_p$ and $(n_n/n_p)_{\mathrm{eq}} = \Gamma_{p \to n}/\Gamma_{n \to p}$. Corrections to the Born rates are shown in Fig.~\ref{fig:detailedbalance}. Equations \eqref{eq:detailedbalance} and thus \eqref{eq:dxnsigma} are not absolutely valid for $\gamma \simeq 1$ because deviations from detailed balance start earlier, but we can nonetheless estimate from this plot that $\sigma_\nu(\gamma \simeq 1) \simeq 0.0008$. With $X_n(\gamma \simeq 1) \simeq 0.2$, we find from Eq.~\eqref{eq:dxnsigma} that including the spectral distortions increases the neutron fraction at freeze-out by
\begin{equation}
\label{eq:shift_disto}
\delta X_n^{[\rm FO],\delta g_{\nu_e}}  \lesssim 0.06 \, \% \, .
\end{equation}
This value is associated with the shift from the orange curve to the green curve in the top panel of Fig.~\ref{fig:analyzevariations}:
\begin{equation}
\delta X_n^{[\rm FO],\delta g_{\nu_e}}  \equiv \delta X_{n,[T_{\nu_e},\text{with dist.}]}^{[\rm FO]} - \delta X_{n,[T_{\nu_e},\text{no dist.}]}^{[\rm FO]} \, .
\end{equation}
The value \eqref{eq:shift_disto} is overestimated because at $\gamma =1$, the neutron-to-proton ratio has already deviated from nuclear statistical equilibrium. In fact, one can reasonably consider that the shift in $\delta X_n^{[\rm FO]}$ is due to the deviation from detailed balance at higher temperatures, when nuclear statistical equilibrium was actually verified (namely, for $\gamma \sim 3$). Indeed, using Eq.~\eqref{eq:dxnsigma} for $\gamma \sim 3$, we obtain the observed shift $\delta X_n^{[\rm FO],\delta g_{\nu_e}} = 0.03 \, \%$.

\vspace{1cm}

We conclude this detailed analysis of neutron/proton freeze-out by stating the obtained value for $\delta X_n^{[\rm FO]}$, which can be read from Fig.~\ref{fig:analyzevariations} at $\gamma \sim 1$ in the ``$T_{\nu_e},$ with distortions" case:
\begin{equation}
\label{eq:dxnfo}
\delta X_n^{[\rm FO]} \simeq - 0.001 \, \% \, .
\end{equation}

\subsubsection{Clock effect}

The {\it clock effect} is due to the higher radiation energy density for a given plasma temperature, which reduces the time necessary to go from $\TFO$ to $\TNuc$. This leads to less neutron beta decay, and thus a higher $X_n(\TNuc)$ and consequently a higher $\YP$. To estimate this contribution we will make several assumptions, justified by observing Fig.~\ref{fig:summary}. Since $t_{\rm Nuc} \sim 245 \, \mathrm{s} \gg t_{\rm FO}$, the freeze-out modification discussed previously will only result in a very small change in duration; indeed, we find numerically that $\Delta t_{\rm FO} \simeq 0.002 \ \mathrm{s}$. We also checked that $\TNuc$ is almost not modified ($\delta \TNuc \simeq - 0.01 \ \%$), which is expected since the onset of nucleosynthesis is essentially determined only by $T_\gamma$. Therefore, the clock effect is mainly described by the change of duration between $\TFO^{(0)}$ and $\TNuc \simeq \TNuc^{(0)}$.

An additional assumption is made by observing the time scale in Fig.~\ref{fig:summary}: most of the neutron beta decay takes place when neutrinos have decoupled and electrons and positrons have annihilated. We will thus consider that between the freeze-out and the beginning of nucleosynthesis, neutrinos are decoupled and $N_{\rm eff} \simeq N_{\rm eff}^{\rm fin}$ is constant.

Therefore, we can write $H \propto 1/2t$ (radiation era). Using the Friedmann equation $H^2 \propto \rho$, we get
\begin{equation}
\frac{\Delta t_{\rm Nuc}}{t_{\rm Nuc}^{(0)}} = - \frac12 \left. \frac{\Delta \rho}{\rho^{(0)}}\right|_{T_\gamma = \TNuc} = - \frac12 \left. \frac{\Delta \rho_\nu}{\rho_\nu^{(0)}}\right|_{\TNuc}  \times \frac{\rho_\nu^{(0)}}{\rho^{(0)}}\,.
\end{equation}
This shift in the neutrino energy density is parametrized by $N_{\rm eff}$, while the ratio of instantaneously decoupled energy densities is, at $\TNuc$, $\rho_\nu^{(0)}/\rho^{(0)} \simeq 0.405$. This gives
\begin{equation}
\frac{\Delta t_{\rm Nuc}}{t_{\rm Nuc}^{(0)}} \simeq - \frac{0.405}{2} \times \frac{\Delta N_{\rm eff}}{3} \simeq -2.3\times 10^{-3} \, ,
\end{equation}
with $\Delta N_{\rm eff} = N_{\rm eff} - 3$ without QED corrections. 

This estimate is actually in very good agreement with the numerical result
\begin{equation}
\left. \frac{\Delta t_{\rm Nuc}}{t_{\rm Nuc}^{(0)}}\right|_{\mathtt{PRIMAT}} \simeq -2.1\times 10^{-3} \, .
\end{equation}
 Hence, the estimate for the clock effect contribution is
\begin{equation}
\label{eq:dxnclock}
\delta X_n^{[\Delta t]} = - \frac{\Delta t_{\rm Nuc}}{t_{\rm Nuc}^{(0)}} \times \frac{t_{\rm Nuc}^{(0)}}{\tau_n} \simeq 0.064 \, \% \, .
\end{equation}

\subsubsection{Helium abundance}

The previous study allows us to estimate the change in the $\He4$ abundance. Since most neutrons are converted into $\He4$, by combining Eqs.~\eqref{eq:dxnfo} and \eqref{eq:dxnclock} (``$T_{\nu_e},$ with distortions" case) we get
\begin{equation}
\delta \YP = \delta X_n^{[\rm Nuc]} = \delta X_n^{[\rm FO]} + \delta X_n^{[\Delta t]} \simeq 0.06 \, \% \, ,
\label{eq:deltaYp}
\end{equation}
which is in quite good agreement with the result in Table~\ref{Table:Corrections}. Our value is slightly overestimated, and we would reach an excellent agreement by instead taking the value\footnote{The apparent going back and forth between $\gamma \sim 0.2$ and $\gamma \sim 3$ in the previous sections emphasizes the limit of the ``instantaneous freeze-out model." To match numerical results, $\delta X_n^{[\rm FO]}$ must be \emph{evaluated} at $\gamma \sim 0.2$, i.e., when Eq.~\eqref{eq:modelFO} starts to be true. Nevertheless, the precise role of distortions is \emph{explained} by the modification of nuclear statistical equilibrium, which is only truly valid until $\gamma \sim 3$.} $\delta X_n^{[\rm FO]} = \delta X_n(\gamma \simeq 0.2)$. Indeed, as mentioned before, the criterion $\gamma \sim 1$ for freeze-out is only a rule of thumb, and it was actually pointed out in Ref.~\cite{Pitrou_2018PhysRept} that the neutron abundance is only affected by beta decay at $T_\gamma \simeq 3.3 \times 10^{9} \, \mathrm{K}$, which corresponds to $\gamma \simeq 0.2$.

The different values of $\delta \YP$ depending on the implementations are very well reproduced: since the energy density is always the same, $\delta X_n^{[\Delta t]}$ remains identical, while the varying $\delta X_n^{[\rm FO]}$ (Fig.~\ref{fig:analyzevariations}) controls $\delta \YP$.

\subsubsection{Other abundances}

We now focus on the other light elements produced during BBN, up to $\Be$. To understand the individual variations of abundances due to incomplete neutrino decoupling, in Table~\ref{Table:LightElements} we separate the final abundances of $\He3$, $\rm T$, $\Be$, and $\Li$.

\begin{table}[h]
	\centering
	\begin{tabular}{|l|cccc|}
  	\hline
  & ${\He3}/{\rm H}$&  ${\rm T}/{\rm H}$ &  ${\Be}/{\rm H}$ & ${\Li}/{\rm H} $ \\
  \hline \hline
  $(i/{\rm H})^{(0),\infty}$ & $1.06 \cdot 10^{-5}$ & $7.84 \cdot 10^{-8}$ & $5.36 \cdot 10^{-10}$ & $2.79 \cdot 10^{-11}$ \\
  $\Delta (i/{\rm H})^\infty$ & $1.3 \cdot 10^{-8}$  & $3.2 \cdot 10^{-10}$ & $- 2.3 \cdot 10^{-12}$ & $1.1 \times 10^{-13}$ \\
   $\delta (i/{\rm H})^\infty$ &  $0.12 \, \%$ & $0.41 \, \%$ &$- 0.43 \, \%$ & $0.40 \, \%$   \\ \hline 
\end{tabular}
	\caption{Neutrino-induced corrections to the primordial production of light elements other than $\He4$ and $\rm D$.
	\label{Table:LightElements}}
\end{table}

\renewcommand{\arraystretch}{1.5}

\begin{table*}[!ht]
	\centering
	\begin{tabular}{|l|cccccc|}
  	\hline 
 BBN framework & $\YP$  & $\delta \YP \,  (\%)$ & ${\rm  D}/{\rm H}  \times 10^5$ &$\delta \left({\rm  D}/{\rm H}\right) \,(\%)$&  ${\left(\He3 + \mathrm{T}\right)}/{\rm  H} \times 10^5$ &  ${\left(\Li + \Be\right)}/{\rm  H} \times 10^{10}$ \\
  \hline \hline
  Inst. decoupling, all corrections & $0.24704$ &  $0$ & $2.450$ &$0$& $1.073$  & $5.694$ \\ \hline
  $\widehat{T}_\nu$ & $0.24709$ &  $0.020$ & $2.459$ &$0.36$& $1.074$  & $5.671$ \\
    $T_{\nu_e},$ no distortions & $0.24699$ & $-0.021$ & $2.458$ & $0.34$ & $1.074$  & $5.669$ \\
  $T_{\nu_e},$ with distortions & $0.24709$ &  $0.019$ & $2.459$ &$0.36$& $1.074$  & $5.671$ \\ \hline
\end{tabular}
	\caption{Light element abundances, including all weak rate corrections and QED corrections to plasma thermodynamics, for various implementations of neutrino-induced corrections. See Table~\ref{Table:Corrections} for results at the Born approximation level.
	\label{Table:Full_Corrections}}
\end{table*}

There are two contributions to the change in the final abundance of an element:
\begin{align}
\delta (i/{\rm H})^{\infty} &= \delta X_i^{\infty} - \delta X_{\rm H}^{\infty}  \nonumber \\
&\simeq \delta X_i^{[\Delta t]} + \delta X_n^{[\rm Nuc]} \, .
\label{eq:abund_light}
\end{align}
The variation of the proton final abundance is directly related to $\delta X_n^{[\rm Nuc]}$ given in Eq.~\eqref{eq:deltaxn}, because an increase of $X_n^{[\rm Nuc]}$ corresponds to a higher neutron-to-proton ratio and/or less beta decay, and thus less protons. On the other hand, the variation of $X_i^{\infty}$ is entirely encapsulated in the clock effect contribution $\delta X_i^{[\Delta t]}$ [it does not depend on $X_n(\TNuc)$ at first order, since all light elements except $\He4$ only appear at trace level]. Indeed, nucleosynthesis consists in elements being produced/destroyed until the reaction rates (which depend only on $T_\gamma$) become too small \cite{SmithBBN}. Because of incomplete neutrino decoupling, a given value of $T_\gamma$ is reached sooner and the nuclear reactions have had less time to be efficient. In other words, there is less time to produce or destroy the different elements.\footnote{This argument does not apply to $\He4$ since it is the most stable light element: for such small variations of the expansion rate, almost all neutrons still end up in $\He4$, so $\YP$ is only affected by $\delta X_n^{[\rm Nuc]}$.}

\begin{figure}[!ht]
	\centering
	\includegraphics[trim=0.2cm 0.4cm 0.3cm 0.5cm, clip, width=8.6cm]{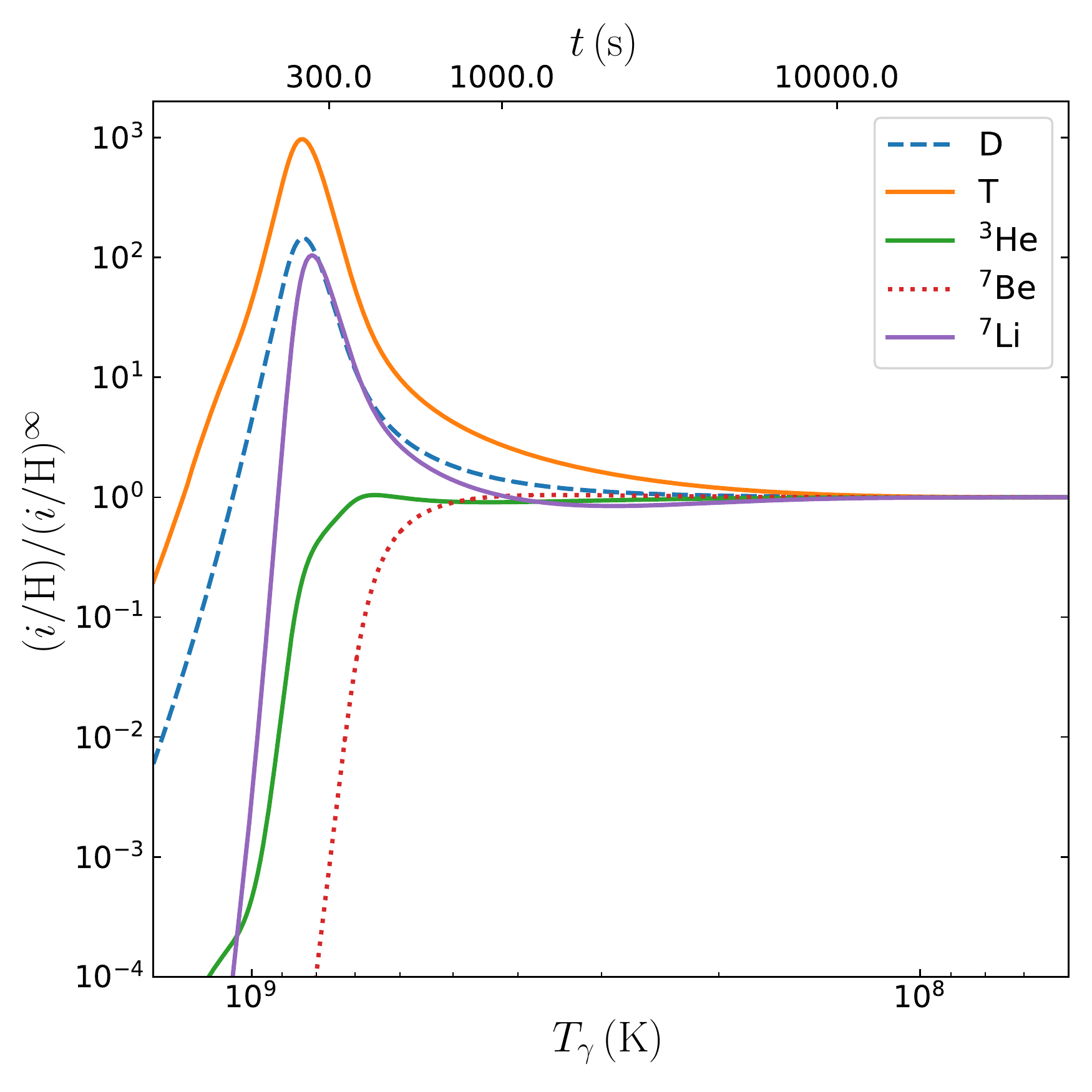}
	\caption{\label{fig:evolutionabundances} Evolution of light element abundances computed with \texttt{PRIMAT}, including incomplete neutrino decoupling at the Born approximation level. To compare the evolutions for different elements, all abundances are rescaled by their frozen-out value.}
\end{figure}

We can thus understand the values of Table~\ref{Table:LightElements} by looking at the evolution of abundances at the end of nucleosynthesis, shown in Fig.~\ref{fig:evolutionabundances}. All elements except $\Be$ are mainly destroyed when the temperature drops below $\TNuc$. The very similar evolutions of $\rm D$, $\rm T$, and $\Li$ explain their similar values of $\delta X_i^\infty$: their destruction rates go to zero more quickly, resulting in a higher final abundance value. For $\Be$ it is the opposite: it is more efficiently produced than destroyed, and the clock effect reduces the possible amount formed (hence, the negative $\delta X_{\Be}^{\infty}$). Moreover, its evolution is even sharper than that of tritium, and thus we expect $\abs{\delta X_{\Be}^{\infty}} > \delta X_{\rm T}^{\infty}$. Finally, $\He3$ has much smaller variations, with a small amplitude of abundance reduction from $\TNuc$. This explains the comparatively small value of $\delta X_{\He3}^{\infty}$.

To recover the aggregated variations of Table~\ref{Table:Previous} (for $\He3$ and $\rm T$, and $\Be$ and $\Li$), one performs the weighted average of individual variations. Since $(\He3/\mathrm{H})^\infty \gg (\mathrm{T}/\mathrm{H})^\infty$, the contribution of $\He3$ dominates, and this argument can be immediately applied to $\Be$ and $\Li$.

\subsection{Precision nucleosynthesis with \texttt{PRIMAT}}
\label{Subsec:other_corrections}

\subsubsection{Full weak rates corrections}

Having thoroughly studied the physics at play by focusing on the Born approximation level, we can now present the results incorporating all weak rates corrections derived in Ref.~\cite{Pitrou_2018PhysRept}. These additional contributions (radiative corrections, finite nucleon mass, and weak magnetism) cannot in principle be added linearly, due to nonlinear feedback between them. Concerning incomplete neutrino decoupling, this means that we also include radiative corrections inside the spectral distortion part of the rates: we modify Eq.~\eqref{eq:deltagamma}, following Eqs.~(100) and (103) in Ref.~\cite{Pitrou_2018PhysRept}.

The results, once again for the three implementations of neutrino-induced corrections, are given in Table~\ref{Table:Full_Corrections}.

Compared to the Born approximation level (Table~\ref{Table:Corrections}), the additional corrections result in higher final abundances, as discussed in Ref.~\cite{Pitrou_2018PhysRept}. Starting then from a baseline where all of these corrections are included except for incomplete neutrino decoupling, the shift in abundances due to neutrinos is slightly reduced by roughly $- \, 0.03 \, \%$; for instance $\delta \YP = + \, 0.02 \, \%$ instead of $+ \, 0.05 \, \%$. The other conclusions of the previous sections remain valid: the average temperature implementation is close to the complete one, we explain $\YP$ through $X_n(\TNuc)$, and the clock effect sources the variations of light elements other than $\He4$.

\begin{figure}[!ht]
	\centering
	\includegraphics[trim=0.1cm 0.4cm 0.2cm 0.2cm, clip, width=8.6cm]{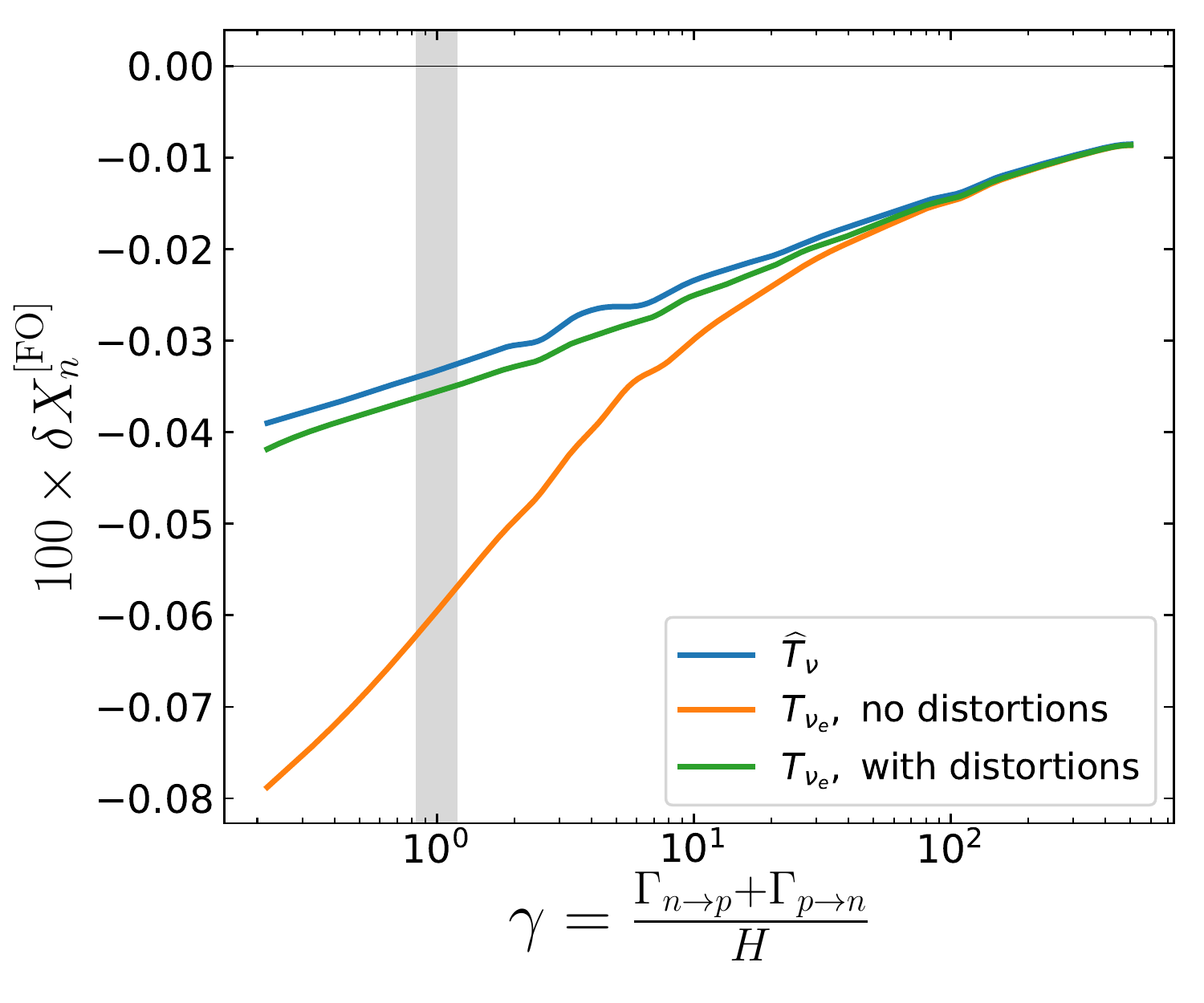}
	\caption{\label{fig:analyzevariationsfull} Neutron fraction around freeze-out, in the different implementations of neutrino-induced corrections. Compared to Fig.~\ref{fig:analyzevariations}, all weak rate corrections are included.}
\end{figure}

Since the additional corrections like finite nucleon mass contributions only affect the weak rates and not the energy density, we expect that the only difference compared to the picture at the Born level will lie in $\delta X_n^{[\rm FO]}$, while $\sigma_\nu$ and $\delta X_i^{[\Delta t]}$ will remain unchanged. This is indeed what we observe in Fig.~\ref{fig:analyzevariationsfull}: the reduction of the neutron fraction at freeze-out due to incomplete neutrino decoupling is enhanced when including all weak rates corrections. Moreover, by comparing Figs.~\ref{fig:analyzevariationsfull} and \ref{fig:analyzevariations} we find
\begin{equation}
\delta X_{n, \mathrm{All}}^{[\rm FO]} - \delta X_{n, \mathrm{Born}}^{[\rm FO]} \simeq - 0.03 \, \% \, ,
\end{equation}
which, by inserting this difference into Eqs.~\eqref{eq:deltaYp} and \eqref{eq:abund_light}, explains the results of Table~\ref{Table:Full_Corrections}.

\subsubsection{What to expect from neutrino oscillations}\label{Sec:Oscillations}

\begin{table*}[!ht]
	\centering
	\begin{tabularx}{\textwidth}{|p{5cm}|XXXX|}
  	\hline 
 BBN framework & $\YP$  & ${\rm  D}/{\rm H}  \times 10^5$ &  ${\left(\He3 + \mathrm{T}\right)}/{\rm  H} \times 10^5$ &  ${\left(\Li + \Be\right)}/{\rm  H} \times 10^{10}$ \\
  \hline \hline
  Inst. decoupling, all corrections & $0.247044$ & $2.45014$ & $1.07276$  & $5.69405$ \\ \hline
  Earlier \texttt{PRIMAT}'s approach ($\widehat{T}_\nu$) & $0.247093$ & $2.45904$ & $1.07404$  & $5.67079$ \\
  Incomplete decoupling ($T_{\nu_e} + \delta g_{\nu_e}$) & $0.247090$ & $2.45903$ &$1.07404$  & $5.67076$ \\ \hline
  Experimental values \cite{Pitrou_2018PhysRept} & $0.2449 \pm 0.0040$ & $2.527 \pm 0.030$ & $< 1.1 \pm 0.2 $ & $1.58^{+ 0.35}_{- 0.28}$ \\ \hline
\end{tabularx}
	\caption{Light element abundances from primordial nucleosynthesis, including all corrections derived in Ref.~\cite{Pitrou_2018PhysRept}. The erroneous implementation of incomplete neutrino decoupling, through an effective average neutrino temperature and no spectral distortions, surprisingly matches the actual results extremely well. We add a digit compared to Table~\ref{Table:Full_Corrections} to highlight the small difference between the two implementations. We recall for comparison the fiducial abundances obtained in the instantaneous decoupling approximation, and the measured values.
	\label{Table:Conclusion}}
\end{table*}

Our findings provide a reasonable guideline for upcoming results that include neutrino oscillations. Progress on this refinement has been made in neutrino evolution calculations \cite{Mangano2005,Relic2016_revisited}, even though some reaction rates (namely, neutrino-neutrino scattering) are still approximate. However, we can forecast the oscillation effect on BBN based on the results of these references. They found that oscillations redistribute the distortions between the different flavors, leaving $N_{\rm eff}$ unchanged, which means that the clock effect contributions will mostly be the same. On the other hand, $z_{\nu_e}$ is reduced (and $z_{\nu_{\mu,\tau}}$ is increased) with smaller $\delta g_{\nu_e}$ distortions, cf.~Fig.~2 in Ref.~\cite{Mangano2005} and Fig.~3 in Ref.~\cite{Relic2016_revisited}. We can thus estimate that, without including the distortions, $\abs{\delta X_n^{[\rm FO]}}$ will be smaller because of $z_{\nu_e}^{\rm osc.} < z_{\nu_e}^{\rm no \, osc.}$. Put differently, the orange curve in Figs.~\ref{fig:analyzevariations} and \ref{fig:analyzevariationsfull} will move closer to the blue one. Then, with smaller distortions the deviation from detailed balance will be reduced ($\sigma_\nu^{\rm osc.}  < \sigma_\nu^{\rm no \, osc.}$ because $||\delta g_{\nu_e}^{\rm osc.}|| < ||\delta g_{\nu_e}^{\rm no \, osc.}||$), and the compensation observed in Figs.~\ref{fig:analyzevariations} and \ref{fig:analyzevariationsfull} should remain, i.e., the green curve will still be close to the blue one. 

In other words, the results from Refs.~\cite{Mangano2005,Relic2016_revisited} indicate that the average temperature implementation should not be modified and would, as in this paper, give results that are remarkably close to the exact implementation. Therefore, we expect that the effects of neutrino oscillations should be subdominant compared to our present discussion.

\section{Conclusion}

In order to assess the consequences of incomplete neutrino decoupling for the production of light elements during BBN, we numerically studied the evolution of neutrino distribution functions through this epoch. Compared to the instantaneous decoupling case, part of the entropy of $e^\pm$ is transferred to the neutrinos, which results in a decrease of the photon comoving temperature and an increased energy density of neutrinos, parametrized by $N_{\rm eff} \simeq 3.044$ when including QED corrections.

We introduced a parametrization of neutrino distribution functions that conveniently separates the energy density change (via effective temperatures) and the remaining spectral distortions. These quantities, obtained throughout the BBN epoch, have been included in the code \texttt{PRIMAT}. The final abundances of light elements, alongside the specific contribution of incomplete neutrino decoupling, are summarized in Table~\ref{Table:Conclusion}. We have been able to scrutinize the physics at play and solve the discrepancy between existing results \cite{Mangano2005,Grohs2015}. The so-called clock effect, due to the increased energy density of neutrinos at a given plasma temperature compared to the fiducial scenario, is responsible for an increase of the deuterium and $\He3$ abundances, and a reduction of the quantity of $\Li$, in agreement with Ref.~\cite{Grohs2015}.

We found that an approximate implementation, assuming that neutrino spectra are purely thermally distorted (``thermal-equivalent distortions" introduced in Refs.~\cite{Dodelson_Turner_PhRvD1992,Fields_PhRvD1993,Mangano2005}), works remarkably well if we set all neutrino species to the same temperature. This puzzling feature is due to a compensation between a delayed neutron/proton freeze-out (because of higher weak rates) and a deviation from detailed balance (because of spectral distortions).

Two additional corrections remain to be included to reach a comprehensive treatment of the physics at play. First, finite-temperature QED corrections to the rates of reactions governing neutrino decoupling need to be computed, but as small corrections to collision terms which are already a correction compared to the fiducial cosmology, these ought to be completely negligible. Then, the introduction of neutrino oscillations needs to use a density matrix formalism and is numerically much more challenging. However, we argued in Sec.~\ref{Sec:Oscillations} that their effect on primordial nucleosynthesis should be subdominant, and thus does not modify our predictions.

\begin{acknowledgments}

J.F. acknowledges financial support through the graduate program of the \'Ecole Normale Sup\'erieure. C.P. and J.F. thank Cristina Volpe for numerous discussions on neutrino physics, and the anonymous referee for his/her constructive comments.

\end{acknowledgments}

\bibliographystyle{apsrev4-1}
\bibliography{Biblio}

\begin{thebibliography}{45}%
\makeatletter
\providecommand \@ifxundefined [1]{%
 \@ifx{#1\undefined}
}%
\providecommand \@ifnum [1]{%
 \ifnum #1\expandafter \@firstoftwo
 \else \expandafter \@secondoftwo
 \fi
}%
\providecommand \@ifx [1]{%
 \ifx #1\expandafter \@firstoftwo
 \else \expandafter \@secondoftwo
 \fi
}%
\providecommand \natexlab [1]{#1}%
\providecommand \enquote  [1]{``#1''}%
\providecommand \bibnamefont  [1]{#1}%
\providecommand \bibfnamefont [1]{#1}%
\providecommand \citenamefont [1]{#1}%
\providecommand \href@noop [0]{\@secondoftwo}%
\providecommand \href [0]{\begingroup \@sanitize@url \@href}%
\providecommand \@href[1]{\@@startlink{#1}\@@href}%
\providecommand \@@href[1]{\endgroup#1\@@endlink}%
\providecommand \@sanitize@url [0]{\catcode `\\12\catcode `\$12\catcode
  `\&12\catcode `\#12\catcode `\^12\catcode `\_12\catcode `\%12\relax}%
\providecommand \@@startlink[1]{}%
\providecommand \@@endlink[0]{}%
\providecommand \url  [0]{\begingroup\@sanitize@url \@url }%
\providecommand \@url [1]{\endgroup\@href {#1}{\urlprefix }}%
\providecommand \urlprefix  [0]{URL }%
\providecommand \Eprint [0]{\href }%
\providecommand \doibase [0]{http://dx.doi.org/}%
\providecommand \selectlanguage [0]{\@gobble}%
\providecommand \bibinfo  [0]{\@secondoftwo}%
\providecommand \bibfield  [0]{\@secondoftwo}%
\providecommand \translation [1]{[#1]}%
\providecommand \BibitemOpen [0]{}%
\providecommand \bibitemStop [0]{}%
\providecommand \bibitemNoStop [0]{.\EOS\space}%
\providecommand \EOS [0]{\spacefactor3000\relax}%
\providecommand \BibitemShut  [1]{\csname bibitem#1\endcsname}%
\let\auto@bib@innerbib\@empty
\bibitem [{\citenamefont {{Izotov}}\ \emph {et~al.}(2014)\citenamefont
  {{Izotov}}, \citenamefont {{Thuan}},\ and\ \citenamefont {{Guseva}}}]{Izo14}%
  \BibitemOpen
  \bibfield  {author} {\bibinfo {author} {\bibfnamefont {Y.~I.}\ \bibnamefont
  {{Izotov}}}, \bibinfo {author} {\bibfnamefont {T.~X.}\ \bibnamefont
  {{Thuan}}}, \ and\ \bibinfo {author} {\bibfnamefont {N.~G.}\ \bibnamefont
  {{Guseva}}},\ }\href {\doibase 10.1093/mnras/stu1771} {\bibfield  {journal}
  {\bibinfo  {journal} {MNRAS}\ }\textbf {\bibinfo {volume} {445}},\ \bibinfo
  {pages} {778} (\bibinfo {year} {2014})},\ \Eprint
  {http://arxiv.org/abs/1408.6953} {arXiv:1408.6953} \BibitemShut {NoStop}%
\bibitem [{\citenamefont {Aver}\ \emph {et~al.}(2015)\citenamefont {Aver},
  \citenamefont {Olive},\ and\ \citenamefont {Skillman}}]{Ave15}%
  \BibitemOpen
  \bibfield  {author} {\bibinfo {author} {\bibfnamefont {E.}~\bibnamefont
  {Aver}}, \bibinfo {author} {\bibfnamefont {K.~A.}\ \bibnamefont {Olive}}, \
  and\ \bibinfo {author} {\bibfnamefont {E.~D.}\ \bibnamefont {Skillman}},\
  }\href {\doibase 10.1088/1475-7516/2015/07/011} {\bibfield  {journal}
  {\bibinfo  {journal} {JCAP}\ }\textbf {\bibinfo {volume} {1507}},\ \bibinfo
  {pages} {011} (\bibinfo {year} {2015})},\ \Eprint
  {http://arxiv.org/abs/1503.08146} {arXiv:1503.08146 [astro-ph.CO]}
  \BibitemShut {NoStop}%
\bibitem [{\citenamefont {{Cooke}}\ \emph {et~al.}(2014)\citenamefont
  {{Cooke}}, \citenamefont {{Pettini}}, \citenamefont {{Jorgenson}},
  \citenamefont {{Murphy}},\ and\ \citenamefont {{Steidel}}}]{Coo14}%
  \BibitemOpen
  \bibfield  {author} {\bibinfo {author} {\bibfnamefont {R.~J.}\ \bibnamefont
  {{Cooke}}}, \bibinfo {author} {\bibfnamefont {M.}~\bibnamefont {{Pettini}}},
  \bibinfo {author} {\bibfnamefont {R.~A.}\ \bibnamefont {{Jorgenson}}},
  \bibinfo {author} {\bibfnamefont {M.~T.}\ \bibnamefont {{Murphy}}}, \ and\
  \bibinfo {author} {\bibfnamefont {C.~C.}\ \bibnamefont {{Steidel}}},\ }\href
  {\doibase 10.1088/0004-637X/781/1/31} {\bibfield  {journal} {\bibinfo
  {journal} {Astrophys. J.}\ }\textbf {\bibinfo {volume} {781}},\ \bibinfo
  {eid} {31} (\bibinfo {year} {2014})},\ \Eprint
  {http://arxiv.org/abs/1308.3240} {arXiv:1308.3240} \BibitemShut {NoStop}%
\bibitem [{\citenamefont {{Cooke}}\ \emph {et~al.}(2016)\citenamefont
  {{Cooke}}, \citenamefont {{Pettini}}, \citenamefont {{Nollett}},\ and\
  \citenamefont {{Jorgenson}}}]{Coo16}%
  \BibitemOpen
  \bibfield  {author} {\bibinfo {author} {\bibfnamefont {R.~J.}\ \bibnamefont
  {{Cooke}}}, \bibinfo {author} {\bibfnamefont {M.}~\bibnamefont {{Pettini}}},
  \bibinfo {author} {\bibfnamefont {K.~M.}\ \bibnamefont {{Nollett}}}, \ and\
  \bibinfo {author} {\bibfnamefont {R.}~\bibnamefont {{Jorgenson}}},\ }\href
  {\doibase 10.3847/0004-637X/830/2/148} {\bibfield  {journal} {\bibinfo
  {journal} {Astrophys. J.}\ }\textbf {\bibinfo {volume} {830}},\ \bibinfo
  {eid} {148} (\bibinfo {year} {2016})},\ \Eprint
  {http://arxiv.org/abs/1607.03900} {arXiv:1607.03900} \BibitemShut {NoStop}%
\bibitem [{\citenamefont {{Cooke}}\ \emph {et~al.}(2018)\citenamefont
  {{Cooke}}, \citenamefont {{Pettini}},\ and\ \citenamefont
  {{Steidel}}}]{Coo18}%
  \BibitemOpen
  \bibfield  {author} {\bibinfo {author} {\bibfnamefont {R.~J.}\ \bibnamefont
  {{Cooke}}}, \bibinfo {author} {\bibfnamefont {M.}~\bibnamefont {{Pettini}}},
  \ and\ \bibinfo {author} {\bibfnamefont {C.~C.}\ \bibnamefont {{Steidel}}},\
  }\href {\doibase 10.3847/1538-4357/aaab53} {\bibfield  {journal} {\bibinfo
  {journal} {\apj}\ }\textbf {\bibinfo {volume} {855}},\ \bibinfo {eid} {102}
  (\bibinfo {year} {2018})},\ \Eprint {http://arxiv.org/abs/1710.11129}
  {arXiv:1710.11129} \BibitemShut {NoStop}%
\bibitem [{\citenamefont {Aghanim}\ \emph {et~al.}(2018)\citenamefont {Aghanim}
  \emph {et~al.}}]{Planck_2018}%
  \BibitemOpen
  \bibfield  {author} {\bibinfo {author} {\bibfnamefont {N.}~\bibnamefont
  {Aghanim}} \emph {et~al.} (\bibinfo {collaboration} {Planck}),\ }\href@noop
  {} {\  (\bibinfo {year} {2018})},\ \Eprint {http://arxiv.org/abs/1807.06209}
  {arXiv:1807.06209 [astro-ph.CO]} \BibitemShut {NoStop}%
\bibitem [{\citenamefont {{Dicus}}\ \emph {et~al.}(1982)\citenamefont
  {{Dicus}}, \citenamefont {{Kolb}}, \citenamefont {{Gleeson}}, \citenamefont
  {{Sudarshan}}, \citenamefont {{Teplitz}},\ and\ \citenamefont
  {{Turner}}}]{Dicus1982}%
  \BibitemOpen
  \bibfield  {author} {\bibinfo {author} {\bibfnamefont {D.~A.}\ \bibnamefont
  {{Dicus}}}, \bibinfo {author} {\bibfnamefont {E.~W.}\ \bibnamefont {{Kolb}}},
  \bibinfo {author} {\bibfnamefont {A.~M.}\ \bibnamefont {{Gleeson}}}, \bibinfo
  {author} {\bibfnamefont {E.~C.~G.}\ \bibnamefont {{Sudarshan}}}, \bibinfo
  {author} {\bibfnamefont {V.~L.}\ \bibnamefont {{Teplitz}}}, \ and\ \bibinfo
  {author} {\bibfnamefont {M.~S.}\ \bibnamefont {{Turner}}},\ }\href {\doibase
  10.1103/PhysRevD.26.2694} {\bibfield  {journal} {\bibinfo  {journal} {Phys.
  Rev.}\ }\textbf {\bibinfo {volume} {D26}},\ \bibinfo {pages} {2694} (\bibinfo
  {year} {1982})}\BibitemShut {NoStop}%
\bibitem [{\citenamefont {Lopez}\ \emph {et~al.}(1997)\citenamefont {Lopez},
  \citenamefont {Turner},\ and\ \citenamefont {Gyuk}}]{Lopez1997}%
  \BibitemOpen
  \bibfield  {author} {\bibinfo {author} {\bibfnamefont {R.~E.}\ \bibnamefont
  {Lopez}}, \bibinfo {author} {\bibfnamefont {M.~S.}\ \bibnamefont {Turner}}, \
  and\ \bibinfo {author} {\bibfnamefont {G.}~\bibnamefont {Gyuk}},\ }\href
  {\doibase 10.1103/PhysRevD.56.3191} {\bibfield  {journal} {\bibinfo
  {journal} {Phys. Rev.}\ }\textbf {\bibinfo {volume} {D56}},\ \bibinfo {pages}
  {3191} (\bibinfo {year} {1997})},\ \Eprint
  {http://arxiv.org/abs/astro-ph/9703065} {arXiv:astro-ph/9703065 [astro-ph]}
  \BibitemShut {NoStop}%
\bibitem [{\citenamefont {{Lopez}}\ and\ \citenamefont
  {{Turner}}(1999)}]{LopezTurner1998}%
  \BibitemOpen
  \bibfield  {author} {\bibinfo {author} {\bibfnamefont {R.~E.}\ \bibnamefont
  {{Lopez}}}\ and\ \bibinfo {author} {\bibfnamefont {M.~S.}\ \bibnamefont
  {{Turner}}},\ }\href {\doibase 10.1103/PhysRevD.59.103502} {\bibfield
  {journal} {\bibinfo  {journal} {Phys. Rev.}\ }\textbf {\bibinfo {volume}
  {D59}},\ \bibinfo {eid} {103502} (\bibinfo {year} {1999})},\ \Eprint
  {http://arxiv.org/abs/astro-ph/9807279} {astro-ph/9807279} \BibitemShut
  {NoStop}%
\bibitem [{\citenamefont {Brown}\ and\ \citenamefont
  {Sawyer}(2001)}]{BrownSawyer}%
  \BibitemOpen
  \bibfield  {author} {\bibinfo {author} {\bibfnamefont {L.~S.}\ \bibnamefont
  {Brown}}\ and\ \bibinfo {author} {\bibfnamefont {R.~F.}\ \bibnamefont
  {Sawyer}},\ }\href {\doibase 10.1103/PhysRevD.63.083503} {\bibfield
  {journal} {\bibinfo  {journal} {Phys. Rev. D}\ }\textbf {\bibinfo {volume}
  {63}},\ \bibinfo {pages} {083503} (\bibinfo {year} {2001})}\BibitemShut
  {NoStop}%
\bibitem [{\citenamefont {Serpico}\ \emph {et~al.}(2004)\citenamefont
  {Serpico}, \citenamefont {Esposito}, \citenamefont {Iocco}, \citenamefont
  {Mangano}, \citenamefont {Miele},\ and\ \citenamefont
  {Pisanti}}]{Serpico:2004gx}%
  \BibitemOpen
  \bibfield  {author} {\bibinfo {author} {\bibfnamefont {P.~D.}\ \bibnamefont
  {Serpico}}, \bibinfo {author} {\bibfnamefont {S.}~\bibnamefont {Esposito}},
  \bibinfo {author} {\bibfnamefont {F.}~\bibnamefont {Iocco}}, \bibinfo
  {author} {\bibfnamefont {G.}~\bibnamefont {Mangano}}, \bibinfo {author}
  {\bibfnamefont {G.}~\bibnamefont {Miele}}, \ and\ \bibinfo {author}
  {\bibfnamefont {O.}~\bibnamefont {Pisanti}},\ }\href {\doibase
  10.1088/1475-7516/2004/12/010} {\bibfield  {journal} {\bibinfo  {journal}
  {JCAP}\ }\textbf {\bibinfo {volume} {0412}},\ \bibinfo {pages} {010}
  (\bibinfo {year} {2004})},\ \Eprint {http://arxiv.org/abs/astro-ph/0408076}
  {arXiv:astro-ph/0408076 [astro-ph]} \BibitemShut {NoStop}%
\bibitem [{\citenamefont {Pitrou}\ \emph {et~al.}(2018)\citenamefont {Pitrou},
  \citenamefont {Coc}, \citenamefont {Uzan},\ and\ \citenamefont
  {Vangioni}}]{Pitrou_2018PhysRept}%
  \BibitemOpen
  \bibfield  {author} {\bibinfo {author} {\bibfnamefont {C.}~\bibnamefont
  {Pitrou}}, \bibinfo {author} {\bibfnamefont {A.}~\bibnamefont {Coc}},
  \bibinfo {author} {\bibfnamefont {J.-P.}\ \bibnamefont {Uzan}}, \ and\
  \bibinfo {author} {\bibfnamefont {E.}~\bibnamefont {Vangioni}},\ }\href
  {\doibase https://doi.org/10.1016/j.physrep.2018.04.005} {\bibfield
  {journal} {\bibinfo  {journal} {Physics Reports}\ }\textbf {\bibinfo {volume}
  {754}},\ \bibinfo {pages} {1 } (\bibinfo {year} {2018})},\ \Eprint
  {http://arxiv.org/abs/1801.08023} {arXiv:1801.08023} \BibitemShut {NoStop}%
\bibitem [{\citenamefont {Pisanti}\ \emph {et~al.}(2008)\citenamefont
  {Pisanti}, \citenamefont {Cirillo}, \citenamefont {Esposito}, \citenamefont
  {Iocco}, \citenamefont {Mangano}, \citenamefont {Miele},\ and\ \citenamefont
  {Serpico}}]{Parthenope}%
  \BibitemOpen
  \bibfield  {author} {\bibinfo {author} {\bibfnamefont {O.}~\bibnamefont
  {Pisanti}}, \bibinfo {author} {\bibfnamefont {A.}~\bibnamefont {Cirillo}},
  \bibinfo {author} {\bibfnamefont {S.}~\bibnamefont {Esposito}}, \bibinfo
  {author} {\bibfnamefont {F.}~\bibnamefont {Iocco}}, \bibinfo {author}
  {\bibfnamefont {G.}~\bibnamefont {Mangano}}, \bibinfo {author} {\bibfnamefont
  {G.}~\bibnamefont {Miele}}, \ and\ \bibinfo {author} {\bibfnamefont
  {P.}~\bibnamefont {Serpico}},\ }\href {\doibase
  https://doi.org/10.1016/j.cpc.2008.02.015} {\bibfield  {journal} {\bibinfo
  {journal} {Computer Physics Communications}\ }\textbf {\bibinfo {volume}
  {178}},\ \bibinfo {pages} {956 } (\bibinfo {year} {2008})}\BibitemShut
  {NoStop}%
\bibitem [{\citenamefont {Consiglio}\ \emph {et~al.}(2018)\citenamefont
  {Consiglio}, \citenamefont {de~Salas}, \citenamefont {Mangano}, \citenamefont
  {Miele}, \citenamefont {Pastor},\ and\ \citenamefont
  {Pisanti}}]{Parthenope_reloaded}%
  \BibitemOpen
  \bibfield  {author} {\bibinfo {author} {\bibfnamefont {R.}~\bibnamefont
  {Consiglio}}, \bibinfo {author} {\bibfnamefont {P.}~\bibnamefont {de~Salas}},
  \bibinfo {author} {\bibfnamefont {G.}~\bibnamefont {Mangano}}, \bibinfo
  {author} {\bibfnamefont {G.}~\bibnamefont {Miele}}, \bibinfo {author}
  {\bibfnamefont {S.}~\bibnamefont {Pastor}}, \ and\ \bibinfo {author}
  {\bibfnamefont {O.}~\bibnamefont {Pisanti}},\ }\href {\doibase
  https://doi.org/10.1016/j.cpc.2018.06.022} {\bibfield  {journal} {\bibinfo
  {journal} {Computer Physics Communications}\ }\textbf {\bibinfo {volume}
  {233}},\ \bibinfo {pages} {237 } (\bibinfo {year} {2018})}\BibitemShut
  {NoStop}%
\bibitem [{\citenamefont {Arbey}(2012)}]{alterbbn2012}%
  \BibitemOpen
  \bibfield  {author} {\bibinfo {author} {\bibfnamefont {A.}~\bibnamefont
  {Arbey}},\ }\href {\doibase https://doi.org/10.1016/j.cpc.2012.03.018}
  {\bibfield  {journal} {\bibinfo  {journal} {Computer Physics Communications}\
  }\textbf {\bibinfo {volume} {183}},\ \bibinfo {pages} {1822 } (\bibinfo
  {year} {2012})}\BibitemShut {NoStop}%
\bibitem [{\citenamefont {Arbey}\ \emph {et~al.}(2020)\citenamefont {Arbey},
  \citenamefont {Auffinger}, \citenamefont {Hickerson},\ and\ \citenamefont
  {Jenssen}}]{alterbbn2018}%
  \BibitemOpen
  \bibfield  {author} {\bibinfo {author} {\bibfnamefont {A.}~\bibnamefont
  {Arbey}}, \bibinfo {author} {\bibfnamefont {J.}~\bibnamefont {Auffinger}},
  \bibinfo {author} {\bibfnamefont {K.}~\bibnamefont {Hickerson}}, \ and\
  \bibinfo {author} {\bibfnamefont {E.}~\bibnamefont {Jenssen}},\ }\href
  {\doibase https://doi.org/10.1016/j.cpc.2019.106982} {\bibfield  {journal}
  {\bibinfo  {journal} {Computer Physics Communications}\ }\textbf {\bibinfo
  {volume} {248}},\ \bibinfo {pages} {106982} (\bibinfo {year}
  {2020})}\BibitemShut {NoStop}%
\bibitem [{\citenamefont {Pitrou}\ and\ \citenamefont
  {Pospelov}(2019)}]{Pitrou:2019pqh}%
  \BibitemOpen
  \bibfield  {author} {\bibinfo {author} {\bibfnamefont {C.}~\bibnamefont
  {Pitrou}}\ and\ \bibinfo {author} {\bibfnamefont {M.}~\bibnamefont
  {Pospelov}},\ }\href@noop {} {\  (\bibinfo {year} {2019})},\ \Eprint
  {http://arxiv.org/abs/1904.07795} {arXiv:1904.07795 [astro-ph.CO]}
  \BibitemShut {NoStop}%
\bibitem [{\citenamefont {Dolgov}\ \emph {et~al.}(1997)\citenamefont {Dolgov},
  \citenamefont {Hansen},\ and\ \citenamefont {Semikoz}}]{Dolgov1997}%
  \BibitemOpen
  \bibfield  {author} {\bibinfo {author} {\bibfnamefont {A.~D.}\ \bibnamefont
  {Dolgov}}, \bibinfo {author} {\bibfnamefont {S.~H.}\ \bibnamefont {Hansen}},
  \ and\ \bibinfo {author} {\bibfnamefont {D.~V.}\ \bibnamefont {Semikoz}},\
  }\href {\doibase 10.1016/S0550-3213(97)00479-3} {\bibfield  {journal}
  {\bibinfo  {journal} {Nuclear Physics B}\ }\textbf {\bibinfo {volume}
  {503}},\ \bibinfo {pages} {426} (\bibinfo {year} {1997})},\ \Eprint
  {http://arxiv.org/abs/hep-ph/9703315} {arXiv:hep-ph/9703315 [hep-ph]}
  \BibitemShut {NoStop}%
\bibitem [{\citenamefont {{Esposito}}\ \emph {et~al.}(2000)\citenamefont
  {{Esposito}}, \citenamefont {{Miele}}, \citenamefont {{Pastor}},
  \citenamefont {{Peloso}},\ and\ \citenamefont
  {{Pisanti}}}]{Esposito_NuPhB2000}%
  \BibitemOpen
  \bibfield  {author} {\bibinfo {author} {\bibfnamefont {S.}~\bibnamefont
  {{Esposito}}}, \bibinfo {author} {\bibfnamefont {G.}~\bibnamefont {{Miele}}},
  \bibinfo {author} {\bibfnamefont {S.}~\bibnamefont {{Pastor}}}, \bibinfo
  {author} {\bibfnamefont {M.}~\bibnamefont {{Peloso}}}, \ and\ \bibinfo
  {author} {\bibfnamefont {O.}~\bibnamefont {{Pisanti}}},\ }\href {\doibase
  10.1016/S0550-3213(00)00554-X} {\bibfield  {journal} {\bibinfo  {journal}
  {Nuclear Physics B}\ }\textbf {\bibinfo {volume} {590}},\ \bibinfo {pages}
  {539} (\bibinfo {year} {2000})},\ \Eprint
  {http://arxiv.org/abs/astro-ph/0005573} {astro-ph/0005573} \BibitemShut
  {NoStop}%
\bibitem [{\citenamefont {{Mangano}}\ \emph {et~al.}(2002)\citenamefont
  {{Mangano}}, \citenamefont {{Miele}}, \citenamefont {{Pastor}},\ and\
  \citenamefont {{Peloso}}}]{Mangano2002}%
  \BibitemOpen
  \bibfield  {author} {\bibinfo {author} {\bibfnamefont {G.}~\bibnamefont
  {{Mangano}}}, \bibinfo {author} {\bibfnamefont {G.}~\bibnamefont {{Miele}}},
  \bibinfo {author} {\bibfnamefont {S.}~\bibnamefont {{Pastor}}}, \ and\
  \bibinfo {author} {\bibfnamefont {M.}~\bibnamefont {{Peloso}}},\ }\href
  {\doibase 10.1016/S0370-2693(02)01622-2} {\bibfield  {journal} {\bibinfo
  {journal} {Physics Letters B}\ }\textbf {\bibinfo {volume} {534}},\ \bibinfo
  {pages} {8} (\bibinfo {year} {2002})},\ \Eprint
  {http://arxiv.org/abs/astro-ph/0111408} {astro-ph/0111408} \BibitemShut
  {NoStop}%
\bibitem [{\citenamefont {Mangano}\ \emph {et~al.}(2005)\citenamefont
  {Mangano}, \citenamefont {Miele}, \citenamefont {Pastor}, \citenamefont
  {Pinto}, \citenamefont {Pisanti},\ and\ \citenamefont
  {Serpico}}]{Mangano2005}%
  \BibitemOpen
  \bibfield  {author} {\bibinfo {author} {\bibfnamefont {G.}~\bibnamefont
  {Mangano}}, \bibinfo {author} {\bibfnamefont {G.}~\bibnamefont {Miele}},
  \bibinfo {author} {\bibfnamefont {S.}~\bibnamefont {Pastor}}, \bibinfo
  {author} {\bibfnamefont {T.}~\bibnamefont {Pinto}}, \bibinfo {author}
  {\bibfnamefont {O.}~\bibnamefont {Pisanti}}, \ and\ \bibinfo {author}
  {\bibfnamefont {P.~D.}\ \bibnamefont {Serpico}},\ }\href {\doibase
  https://doi.org/10.1016/j.nuclphysb.2005.09.041} {\bibfield  {journal}
  {\bibinfo  {journal} {Nuclear Physics B}\ }\textbf {\bibinfo {volume}
  {729}},\ \bibinfo {pages} {221 } (\bibinfo {year} {2005})}\BibitemShut
  {NoStop}%
\bibitem [{\citenamefont {Grohs}\ \emph {et~al.}(2016)\citenamefont {Grohs},
  \citenamefont {Fuller}, \citenamefont {Kishimoto}, \citenamefont {Paris},\
  and\ \citenamefont {Vlasenko}}]{Grohs2015}%
  \BibitemOpen
  \bibfield  {author} {\bibinfo {author} {\bibfnamefont {E.}~\bibnamefont
  {Grohs}}, \bibinfo {author} {\bibfnamefont {G.~M.}\ \bibnamefont {Fuller}},
  \bibinfo {author} {\bibfnamefont {C.~T.}\ \bibnamefont {Kishimoto}}, \bibinfo
  {author} {\bibfnamefont {M.~W.}\ \bibnamefont {Paris}}, \ and\ \bibinfo
  {author} {\bibfnamefont {A.}~\bibnamefont {Vlasenko}},\ }\href {\doibase
  10.1103/PhysRevD.93.083522} {\bibfield  {journal} {\bibinfo  {journal} {Phys.
  Rev. D}\ }\textbf {\bibinfo {volume} {93}},\ \bibinfo {pages} {083522}
  (\bibinfo {year} {2016})}\BibitemShut {NoStop}%
\bibitem [{\citenamefont {Escudero}(2019)}]{Escudero2019}%
  \BibitemOpen
  \bibfield  {author} {\bibinfo {author} {\bibfnamefont {M.}~\bibnamefont
  {Escudero}},\ }\href {\doibase 10.1088/1475-7516/2019/02/007} {\bibfield
  {journal} {\bibinfo  {journal} {JCAP}\ }\textbf {\bibinfo {volume} {1902}},\
  \bibinfo {pages} {007} (\bibinfo {year} {2019})},\ \Eprint
  {http://arxiv.org/abs/1812.05605} {arXiv:1812.05605 [hep-ph]} \BibitemShut
  {NoStop}%
\bibitem [{\citenamefont {{Hannestad}}\ and\ \citenamefont
  {{Madsen}}(1995)}]{Hannestad_PhRvD1995}%
  \BibitemOpen
  \bibfield  {author} {\bibinfo {author} {\bibfnamefont {S.}~\bibnamefont
  {{Hannestad}}}\ and\ \bibinfo {author} {\bibfnamefont {J.}~\bibnamefont
  {{Madsen}}},\ }\href {\doibase 10.1103/PhysRevD.52.1764} {\bibfield
  {journal} {\bibinfo  {journal} {\prd}\ }\textbf {\bibinfo {volume} {52}},\
  \bibinfo {pages} {1764} (\bibinfo {year} {1995})},\ \Eprint
  {http://arxiv.org/abs/astro-ph/9506015} {astro-ph/9506015} \BibitemShut
  {NoStop}%
\bibitem [{\citenamefont {{Thomas}}\ \emph {et~al.}(2019)\citenamefont
  {{Thomas}}, \citenamefont {{Dezen}}, \citenamefont {{Grohs}},\ and\
  \citenamefont {{Kishimoto}}}]{Grohs2019}%
  \BibitemOpen
  \bibfield  {author} {\bibinfo {author} {\bibfnamefont {L.~C.}\ \bibnamefont
  {{Thomas}}}, \bibinfo {author} {\bibfnamefont {T.}~\bibnamefont {{Dezen}}},
  \bibinfo {author} {\bibfnamefont {E.~B.}\ \bibnamefont {{Grohs}}}, \ and\
  \bibinfo {author} {\bibfnamefont {C.~T.}\ \bibnamefont {{Kishimoto}}},\
  }\href@noop {} {\  (\bibinfo {year} {2019})},\ \Eprint
  {http://arxiv.org/abs/1910.14050} {arXiv:1910.14050 [hep-ph]} \BibitemShut
  {NoStop}%
\bibitem [{\citenamefont {{Heckler}}(1994)}]{Heckler_PhRvD1994}%
  \BibitemOpen
  \bibfield  {author} {\bibinfo {author} {\bibfnamefont {A.~F.}\ \bibnamefont
  {{Heckler}}},\ }\href {\doibase 10.1103/PhysRevD.49.611} {\bibfield
  {journal} {\bibinfo  {journal} {\prd}\ }\textbf {\bibinfo {volume} {49}},\
  \bibinfo {pages} {611} (\bibinfo {year} {1994})}\BibitemShut {NoStop}%
\bibitem [{\citenamefont {{Fornengo}}\ \emph {et~al.}(1997)\citenamefont
  {{Fornengo}}, \citenamefont {{Kim}},\ and\ \citenamefont
  {{Song}}}]{Fornengo_PhRvD1997}%
  \BibitemOpen
  \bibfield  {author} {\bibinfo {author} {\bibfnamefont {N.}~\bibnamefont
  {{Fornengo}}}, \bibinfo {author} {\bibfnamefont {C.~W.}\ \bibnamefont
  {{Kim}}}, \ and\ \bibinfo {author} {\bibfnamefont {J.}~\bibnamefont
  {{Song}}},\ }\href {\doibase 10.1103/PhysRevD.56.5123} {\bibfield  {journal}
  {\bibinfo  {journal} {\prd}\ }\textbf {\bibinfo {volume} {56}},\ \bibinfo
  {pages} {5123} (\bibinfo {year} {1997})},\ \Eprint
  {http://arxiv.org/abs/hep-ph/9702324} {hep-ph/9702324} \BibitemShut {NoStop}%
\bibitem [{\citenamefont {{Bennett}}\ \emph {et~al.}(2019)\citenamefont
  {{Bennett}}, \citenamefont {{Buldgen}}, \citenamefont {{Drewes}},\ and\
  \citenamefont {{Wong}}}]{Bennett2019}%
  \BibitemOpen
  \bibfield  {author} {\bibinfo {author} {\bibfnamefont {J.~J.}\ \bibnamefont
  {{Bennett}}}, \bibinfo {author} {\bibfnamefont {G.}~\bibnamefont
  {{Buldgen}}}, \bibinfo {author} {\bibfnamefont {M.}~\bibnamefont {{Drewes}}},
  \ and\ \bibinfo {author} {\bibfnamefont {Y.~Y.~Y.}\ \bibnamefont {{Wong}}},\
  }\href@noop {} {\bibfield  {journal} {\bibinfo  {journal} {arXiv e-prints}\
  ,\ \bibinfo {eid} {arXiv:1911.04504}} (\bibinfo {year} {2019})},\ \Eprint
  {http://arxiv.org/abs/1911.04504} {arXiv:1911.04504 [hep-ph]} \BibitemShut
  {NoStop}%
\bibitem [{\citenamefont {{Sigl}}\ and\ \citenamefont
  {{Raffelt}}(1993)}]{SiglRaffelt}%
  \BibitemOpen
  \bibfield  {author} {\bibinfo {author} {\bibfnamefont {G.}~\bibnamefont
  {{Sigl}}}\ and\ \bibinfo {author} {\bibfnamefont {G.}~\bibnamefont
  {{Raffelt}}},\ }\href {\doibase 10.1016/0550-3213(93)90175-O} {\bibfield
  {journal} {\bibinfo  {journal} {Nuclear Physics B}\ }\textbf {\bibinfo
  {volume} {406}},\ \bibinfo {pages} {423} (\bibinfo {year}
  {1993})}\BibitemShut {NoStop}%
\bibitem [{\citenamefont {Stirner}\ \emph {et~al.}(2018)\citenamefont
  {Stirner}, \citenamefont {Sigl},\ and\ \citenamefont
  {Raffelt}}]{Stirner2018}%
  \BibitemOpen
  \bibfield  {author} {\bibinfo {author} {\bibfnamefont {T.}~\bibnamefont
  {Stirner}}, \bibinfo {author} {\bibfnamefont {G.}~\bibnamefont {Sigl}}, \
  and\ \bibinfo {author} {\bibfnamefont {G.}~\bibnamefont {Raffelt}},\ }\href
  {\doibase 10.1088/1475-7516/2018/05/016} {\bibfield  {journal} {\bibinfo
  {journal} {JCAP}\ }\textbf {\bibinfo {volume} {1805}},\ \bibinfo {pages}
  {016} (\bibinfo {year} {2018})},\ \Eprint {http://arxiv.org/abs/1803.04693}
  {arXiv:1803.04693 [hep-ph]} \BibitemShut {NoStop}%
\bibitem [{\citenamefont {{Volpe}}\ \emph {et~al.}(2013)\citenamefont
  {{Volpe}}, \citenamefont {{V{\"a}{\"a}n{\"a}nen}},\ and\ \citenamefont
  {{Espinoza}}}]{Volpe_2013}%
  \BibitemOpen
  \bibfield  {author} {\bibinfo {author} {\bibfnamefont {C.}~\bibnamefont
  {{Volpe}}}, \bibinfo {author} {\bibfnamefont {D.}~\bibnamefont
  {{V{\"a}{\"a}n{\"a}nen}}}, \ and\ \bibinfo {author} {\bibfnamefont
  {C.}~\bibnamefont {{Espinoza}}},\ }\href {\doibase
  10.1103/PhysRevD.87.113010} {\bibfield  {journal} {\bibinfo  {journal}
  {\prd}\ }\textbf {\bibinfo {volume} {87}},\ \bibinfo {eid} {113010} (\bibinfo
  {year} {2013})},\ \Eprint {http://arxiv.org/abs/1302.2374} {arXiv:1302.2374
  [hep-ph]} \BibitemShut {NoStop}%
\bibitem [{\citenamefont {{Volpe}}(2015)}]{Volpe_2015}%
  \BibitemOpen
  \bibfield  {author} {\bibinfo {author} {\bibfnamefont {C.}~\bibnamefont
  {{Volpe}}},\ }\href {\doibase 10.1142/S0218301315410098} {\bibfield
  {journal} {\bibinfo  {journal} {International Journal of Modern Physics E}\
  }\textbf {\bibinfo {volume} {24}},\ \bibinfo {eid} {1541009} (\bibinfo {year}
  {2015})},\ \Eprint {http://arxiv.org/abs/1506.06222} {arXiv:1506.06222
  [astro-ph.SR]} \BibitemShut {NoStop}%
\bibitem [{\citenamefont {Vlasenko}\ \emph {et~al.}(2014)\citenamefont
  {Vlasenko}, \citenamefont {Fuller},\ and\ \citenamefont
  {Cirigliano}}]{Vlasenko_PhRevD2014}%
  \BibitemOpen
  \bibfield  {author} {\bibinfo {author} {\bibfnamefont {A.}~\bibnamefont
  {Vlasenko}}, \bibinfo {author} {\bibfnamefont {G.~M.}\ \bibnamefont
  {Fuller}}, \ and\ \bibinfo {author} {\bibfnamefont {V.}~\bibnamefont
  {Cirigliano}},\ }\href {\doibase 10.1103/PhysRevD.89.105004} {\bibfield
  {journal} {\bibinfo  {journal} {Phys. Rev. D}\ }\textbf {\bibinfo {volume}
  {89}},\ \bibinfo {pages} {105004} (\bibinfo {year} {2014})}\BibitemShut
  {NoStop}%
\bibitem [{\citenamefont {Blaschke}\ and\ \citenamefont
  {Cirigliano}(2016)}]{BlaschkeCirigliano}%
  \BibitemOpen
  \bibfield  {author} {\bibinfo {author} {\bibfnamefont {D.~N.}\ \bibnamefont
  {Blaschke}}\ and\ \bibinfo {author} {\bibfnamefont {V.}~\bibnamefont
  {Cirigliano}},\ }\href {\doibase 10.1103/PhysRevD.94.033009} {\bibfield
  {journal} {\bibinfo  {journal} {Phys. Rev. D}\ }\textbf {\bibinfo {volume}
  {94}},\ \bibinfo {pages} {033009} (\bibinfo {year} {2016})}\BibitemShut
  {NoStop}%
\bibitem [{\citenamefont {de~Salas}\ and\ \citenamefont
  {Pastor}(2016)}]{Relic2016_revisited}%
  \BibitemOpen
  \bibfield  {author} {\bibinfo {author} {\bibfnamefont {P.~F.}\ \bibnamefont
  {de~Salas}}\ and\ \bibinfo {author} {\bibfnamefont {S.}~\bibnamefont
  {Pastor}},\ }\href {\doibase 10.1088/1475-7516/2016/07/051} {\bibfield
  {journal} {\bibinfo  {journal} {JCAP}\ }\textbf {\bibinfo {volume} {1607}},\
  \bibinfo {pages} {051} (\bibinfo {year} {2016})},\ \Eprint
  {http://arxiv.org/abs/1606.06986} {arXiv:1606.06986 [hep-ph]} \BibitemShut
  {NoStop}%
\bibitem [{\citenamefont {{Dolgov}}\ \emph {et~al.}(1999)\citenamefont
  {{Dolgov}}, \citenamefont {{Hansen}},\ and\ \citenamefont
  {{Semikoz}}}]{Dolgov_NuPhB1999}%
  \BibitemOpen
  \bibfield  {author} {\bibinfo {author} {\bibfnamefont {A.~D.}\ \bibnamefont
  {{Dolgov}}}, \bibinfo {author} {\bibfnamefont {S.~H.}\ \bibnamefont
  {{Hansen}}}, \ and\ \bibinfo {author} {\bibfnamefont {D.~V.}\ \bibnamefont
  {{Semikoz}}},\ }\href {\doibase 10.1016/S0550-3213(98)00818-9} {\bibfield
  {journal} {\bibinfo  {journal} {Nuclear Physics B}\ }\textbf {\bibinfo
  {volume} {543}},\ \bibinfo {pages} {269} (\bibinfo {year} {1999})},\ \Eprint
  {http://arxiv.org/abs/hep-ph/9805467} {hep-ph/9805467} \BibitemShut {NoStop}%
\bibitem [{\citenamefont {Grohs}\ and\ \citenamefont
  {Fuller}(2017)}]{Grohs_insights}%
  \BibitemOpen
  \bibfield  {author} {\bibinfo {author} {\bibfnamefont {E.}~\bibnamefont
  {Grohs}}\ and\ \bibinfo {author} {\bibfnamefont {G.~M.}\ \bibnamefont
  {Fuller}},\ }\href {\doibase https://doi.org/10.1016/j.nuclphysb.2017.07.019}
  {\bibfield  {journal} {\bibinfo  {journal} {Nuclear Physics B}\ }\textbf
  {\bibinfo {volume} {923}},\ \bibinfo {pages} {222 } (\bibinfo {year}
  {2017})}\BibitemShut {NoStop}%
\bibitem [{\citenamefont {Lucca}\ \emph {et~al.}(2019)\citenamefont {Lucca},
  \citenamefont {Schöneberg}, \citenamefont {Hooper}, \citenamefont
  {Lesgourgues},\ and\ \citenamefont {Chluba}}]{Lucca:2019rxf}%
  \BibitemOpen
  \bibfield  {author} {\bibinfo {author} {\bibfnamefont {M.}~\bibnamefont
  {Lucca}}, \bibinfo {author} {\bibfnamefont {N.}~\bibnamefont {Schöneberg}},
  \bibinfo {author} {\bibfnamefont {D.~C.}\ \bibnamefont {Hooper}}, \bibinfo
  {author} {\bibfnamefont {J.}~\bibnamefont {Lesgourgues}}, \ and\ \bibinfo
  {author} {\bibfnamefont {J.}~\bibnamefont {Chluba}},\ }\href@noop {} {\
  (\bibinfo {year} {2019})},\ \Eprint {http://arxiv.org/abs/1910.04619}
  {arXiv:1910.04619 [astro-ph.CO]} \BibitemShut {NoStop}%
\bibitem [{\citenamefont {Peter}\ and\ \citenamefont
  {Uzan}(2013)}]{Peter_Uzan}%
  \BibitemOpen
  \bibfield  {author} {\bibinfo {author} {\bibfnamefont {P.}~\bibnamefont
  {Peter}}\ and\ \bibinfo {author} {\bibfnamefont {J.-P.}\ \bibnamefont
  {Uzan}},\ }\href@noop {} {\emph {\bibinfo {title} {{Primordial
  Cosmology}}}},\ Oxford Graduate Texts\ (\bibinfo  {publisher} {Oxford
  University Press},\ \bibinfo {year} {2013})\BibitemShut {NoStop}%
\bibitem [{\citenamefont {Lesgourgues}\ \emph {et~al.}(2013)\citenamefont
  {Lesgourgues}, \citenamefont {Mangano}, \citenamefont {Miele},\ and\
  \citenamefont {Pastor}}]{Neutrino_Cosmology}%
  \BibitemOpen
  \bibfield  {author} {\bibinfo {author} {\bibfnamefont {J.}~\bibnamefont
  {Lesgourgues}}, \bibinfo {author} {\bibfnamefont {G.}~\bibnamefont
  {Mangano}}, \bibinfo {author} {\bibfnamefont {G.}~\bibnamefont {Miele}}, \
  and\ \bibinfo {author} {\bibfnamefont {S.}~\bibnamefont {Pastor}},\ }\href
  {https://www.cambridge.org/academic/subjects/physics/particle-physics-and-nuclear-physics/neutrino-cosmology?format=PB&isbn=9781108705011}
  {\emph {\bibinfo {title} {{Neutrino Cosmology}}}}\ (\bibinfo  {publisher}
  {Cambridge University Press},\ \bibinfo {year} {2013})\BibitemShut {NoStop}%
\bibitem [{\citenamefont {Bernstein}\ \emph {et~al.}(1989)\citenamefont
  {Bernstein}, \citenamefont {Brown},\ and\ \citenamefont
  {Feinberg}}]{bernstein1989}%
  \BibitemOpen
  \bibfield  {author} {\bibinfo {author} {\bibfnamefont {J.}~\bibnamefont
  {Bernstein}}, \bibinfo {author} {\bibfnamefont {L.~S.}\ \bibnamefont
  {Brown}}, \ and\ \bibinfo {author} {\bibfnamefont {G.}~\bibnamefont
  {Feinberg}},\ }\href {\doibase 10.1103/RevModPhys.61.25} {\bibfield
  {journal} {\bibinfo  {journal} {Rev. Mod. Phys.}\ }\textbf {\bibinfo {volume}
  {61}},\ \bibinfo {pages} {25} (\bibinfo {year} {1989})}\BibitemShut {NoStop}%
\bibitem [{\citenamefont {{Dodelson}}\ and\ \citenamefont
  {{Turner}}(1992)}]{Dodelson_Turner_PhRvD1992}%
  \BibitemOpen
  \bibfield  {author} {\bibinfo {author} {\bibfnamefont {S.}~\bibnamefont
  {{Dodelson}}}\ and\ \bibinfo {author} {\bibfnamefont {M.~S.}\ \bibnamefont
  {{Turner}}},\ }\href {\doibase 10.1103/PhysRevD.46.3372} {\bibfield
  {journal} {\bibinfo  {journal} {\prd}\ }\textbf {\bibinfo {volume} {46}},\
  \bibinfo {pages} {3372} (\bibinfo {year} {1992})}\BibitemShut {NoStop}%
\bibitem [{\citenamefont {{Fields}}\ \emph {et~al.}(1993)\citenamefont
  {{Fields}}, \citenamefont {{Dodelson}},\ and\ \citenamefont
  {{Turner}}}]{Fields_PhRvD1993}%
  \BibitemOpen
  \bibfield  {author} {\bibinfo {author} {\bibfnamefont {B.~D.}\ \bibnamefont
  {{Fields}}}, \bibinfo {author} {\bibfnamefont {S.}~\bibnamefont
  {{Dodelson}}}, \ and\ \bibinfo {author} {\bibfnamefont {M.~S.}\ \bibnamefont
  {{Turner}}},\ }\href {\doibase 10.1103/PhysRevD.47.4309} {\bibfield
  {journal} {\bibinfo  {journal} {\prd}\ }\textbf {\bibinfo {volume} {47}},\
  \bibinfo {pages} {4309} (\bibinfo {year} {1993})},\ \Eprint
  {http://arxiv.org/abs/astro-ph/9210007} {astro-ph/9210007} \BibitemShut
  {NoStop}%
\bibitem [{\citenamefont {Smith}\ \emph {et~al.}(1993)\citenamefont {Smith},
  \citenamefont {Kawano},\ and\ \citenamefont {Malaney}}]{SmithBBN}%
  \BibitemOpen
  \bibfield  {author} {\bibinfo {author} {\bibfnamefont {M.~S.}\ \bibnamefont
  {Smith}}, \bibinfo {author} {\bibfnamefont {L.~H.}\ \bibnamefont {Kawano}}, \
  and\ \bibinfo {author} {\bibfnamefont {R.~A.}\ \bibnamefont {Malaney}},\
  }\href {\doibase 10.1086/191763} {\bibfield  {journal} {\bibinfo  {journal}
  {Astrophys. J. Suppl.}\ }\textbf {\bibinfo {volume} {85}},\ \bibinfo {pages}
  {219} (\bibinfo {year} {1993})}\BibitemShut {NoStop}%
\bibitem [{\citenamefont {{Semikoz}}\ and\ \citenamefont
  {{Tkachev}}(1997)}]{Semikoz_1997PhRvD}%
  \BibitemOpen
  \bibfield  {author} {\bibinfo {author} {\bibfnamefont {D.~V.}\ \bibnamefont
  {{Semikoz}}}\ and\ \bibinfo {author} {\bibfnamefont {I.~I.}\ \bibnamefont
  {{Tkachev}}},\ }\href {\doibase 10.1103/PhysRevD.55.489} {\bibfield
  {journal} {\bibinfo  {journal} {\prd}\ }\textbf {\bibinfo {volume} {55}},\
  \bibinfo {pages} {489} (\bibinfo {year} {1997})},\ \Eprint
  {http://arxiv.org/abs/hep-ph/9507306} {hep-ph/9507306} \BibitemShut {NoStop}%
\end{thebibliography}%

\appendix

\section{Neutrino transport equations}
\label{App:equations}

Neutrino evolution is computed by simultaneously solving a total of nine equations. The first eight correspond to rewriting the Boltzmann equations \eqref{eq:da_dx} for $i=0,\dots,4$ and $\alpha=e,\mu$. The collision integrals appearing on the right-hand side are reduced to two-dimensional integrals following the method outlined in Refs.~\cite{Dolgov1997,Semikoz_1997PhRvD}, and read
\begin{widetext}
\begin{align}
\begin{split}
 	C_{\nu_e}(x,y_1) = &\frac{m_e^5 G_F^2}{2 \pi^3 y_1 x^5} \int{\dd y_2 \, y_2 \, \dd y_3 \, y_3 \, \dd y_4 \, y_4 \, \delta(E_1+E_2-E_3-E_4)} \\
 	&\times \Big\{ F[f_{\nu_e}^{(1)},f_{\nu_e}^{(2)},f_{\nu_e}^{(3)},f_{\nu_e}^{(4)}] \left(6 d_1 - 4 d_2(1,4) - 4 d_2(2,3) + 2 d_2(1,2) + 2 d_2(3,4) + 6 d_3 \right) \\
	&+ F[f_{\nu_e}^{(1)},f_{\nu_\mu}^{(2)},f_{\nu_e}^{(3)},f_{\nu_\mu}^{(4)}] \left(4 d_1 - 2 d_2(1,4) - 2 d_2(2,3) + 2 d_2(1,2) + 2 d_2(3,4) + 4 d_3 \right) \\
	&+ F[f_{\nu_e}^{(1)},f_{\nu_e}^{(2)},f_{\nu_\mu}^{(3)},f_{\nu_\mu}^{(4)}] \left(2 d_1 - 2 d_2(1,4) - 2 d_2(2,3) + 2 d_3 \right) \\
	&+ F[f_{\nu_e}^{(1)},f_{e}^{(2)},f_{\nu_e}^{(3)},f_{e}^{(4)}] \left[4(g_L^2 +g_R^2) \left(2d_1 - d_2(1,4) - d_2(2,3) + d_2(1,2) + d_2(3,4) + 2 d_3\right) \right. \\
	&\phantom{+ F[f_{\nu_e}^{(1)},f_{e}^{(2)},f_{\nu_e}^{(3)},f_{e}^{(4)}]} \qquad \qquad \qquad \qquad \left. - 8 g_L g_R x^2\left(d_1 - d_2(1,3)\right)/E_2 E_4 \right] \\
	&+ F[f_{\nu_e}^{(1)},f_{\nu_e}^{(2)},f_{e}^{(3)},f_{e}^{(4)}] \left[4 g_R^2 \left(d_1 - d_2(1,4) - d_2(2,3) + d_3\right) + 4 g_L^2 \left(d_1 - d_2(2,4) - d_2(1,3) + d_3\right) \right. \\
	&\phantom{+ F[f_{\nu_e}^{(1)},f_{e}^{(2)},f_{\nu_e}^{(3)},f_{e}^{(4)}]} \qquad \qquad \qquad \qquad \left. + 4 g_L g_R x^2\left(d_1 + d_2(1,2)\right)/E_3 E_4 \right] \Big\} \, ,
\end{split} \label{eq:cnu_e} \\ \nonumber \\\nonumber \\ \nonumber\\
\begin{split}
 	C_{\nu_\mu}(x,y_1) = &\frac{m_e^5 G_F^2}{2 \pi^3 y_1 x^5} \int{\dd y_2 \, y_2 \, \dd y_3 \, y_3 \, \dd y_4 \, y_4 \, \delta(E_1+E_2-E_3-E_4)} \\
 	&\times \Big\{ F[f_{\nu_\mu}^{(1)},f_{\nu_\mu}^{(2)},f_{\nu_\mu}^{(3)},f_{\nu_\mu}^{(4)}] \left(9 d_1 - 6 d_2(1,4) - 6 d_2(2,3) + 3 d_2(1,2) + 3 d_2(3,4) + 9 d_3 \right) \\
	&+ F[f_{\nu_\mu}^{(1)},f_{\nu_e}^{(2)},f_{\nu_\mu}^{(3)},f_{\nu_e}^{(4)}] \left(2 d_1 - d_2(1,4) - d_2(2,3) + d_2(1,2) + d_2(3,4) + 2 d_3 \right) \\
	&+ F[f_{\nu_\mu}^{(1)},f_{\nu_\mu}^{(2)},f_{\nu_e}^{(3)},f_{\nu_e}^{(4)}] \left(d_1 - d_2(1,4) - d_2(2,3) + d_3 \right) \\
	&+ F[f_{\nu_\mu}^{(1)},f_{e}^{(2)},f_{\nu_\mu}^{(3)},f_{e}^{(4)}] \left[4(\tilde{g}_L^2 +g_R^2) \left(2d_1 - d_2(1,4) - d_2(2,3) + d_2(1,2) + d_2(3,4) + 2 d_3\right) \right. \\
	&\phantom{+ F[f_{\nu_e}^{(1)},f_{e}^{(2)},f_{\nu_e}^{(3)},f_{e}^{(4)}]} \qquad \qquad \qquad \qquad \left. - 8 \tilde{g}_L g_R x^2\left(d_1 - d_2(1,3)\right)/E_2 E_4 \right] \\
	&+ F[f_{\nu_\mu}^{(1)},f_{\nu_\mu}^{(2)},f_{e}^{(3)},f_{e}^{(4)}] \left[4 g_R^2 \left(d_1 - d_2(1,4) - d_2(2,3) + d_3\right) + 4 \tilde{g}_L^2 \left(d_1 - d_2(2,4) - d_2(1,3) + d_3\right) \right. \\
	&\phantom{+ F[f_{\nu_e}^{(1)},f_{e}^{(2)},f_{\nu_e}^{(3)},f_{e}^{(4)}]} \qquad \qquad \qquad \qquad \left. + 4 \tilde{g}_L g_R x^2\left(d_1 + d_2(1,2)\right)/E_3 E_4 \right] \Big\} \, .
\end{split} \label{eq:cnu_mu}
\end{align}
\end{widetext}
To standardize the notations, we wrote $d_1 \equiv D_1$, $d_2(i,j) \equiv D_2(i,j)/E_iE_j$, $d_3 = D_3/E_1E_2E_3E_4$, and \[F \equiv f^{(3)} f^{(4)}(1-f^{(1)})(1-f^{(2)}) - f^{(1)} f^{(2)} (1-f^{(3)})(1-f^{(4)}) \, ,\]
where $f_a^{(j)}$ denotes $f_a(y_j)$. The functions $D_j$ are defined in Ref.~\cite{Dolgov1997}. The weak interaction couplings are $\tilde{g}_L = \sin^2{\theta_W} + 1/2$ for $\nu_e$, $g_L = \sin^2{\theta_W} - 1/2$ for $\nu_{\mu,\tau}$, and $g_R = \sin^2{\theta_W}$ for all species. This difference between flavors is due to charged-current processes; its consequences were discussed in Sec.~\ref{subsec:neutrino_results}.  Finally, some typos were corrected compared to the corresponding Eqs.~(9)-(10) in Ref.~\cite{Dolgov1997}.

The last equation describes the evolution of the plasma temperature [cf. Eq.~(15) of Ref.~\cite{Esposito_NuPhB2000}],
\begin{equation}
\frac{dz}{dx} =  \frac{\displaystyle \frac{x}{z}J(x/z) - \frac{1}{2 \pi^2 z^3} \frac{1}{xH} \int_{0}^{\infty}{\dd y \, y^3 \left(C_{\nu_e} + 2C_{\nu_\mu}\right)}}{ \displaystyle \frac{x^2}{z^2}J(x/z) + Y(x/z) + \frac{2 \pi^2}{15}} \, , \label{eq:z}
\end{equation}
where we introduced
\begin{align}
J(\tau) &\equiv \frac{1}{\pi^2} \int_{0}^{\infty}{\dd \omega \, \omega^2 \frac{\exp{(\sqrt{\omega^2 + \tau^2})}}{\left[\exp{(\sqrt{\omega^2 + \tau^2})}+1\right]^2}} \, , \\
Y(\tau) &\equiv \frac{1}{\pi^2} \int_{0}^{\infty}{\dd \omega \, \omega^4 \frac{\exp{(\sqrt{\omega^2 + \tau^2})}}{\left[\exp{(\sqrt{\omega^2 + \tau^2})}+1\right]^2}} \, .
\end{align}
This equation is derived by rewriting the continuity equation in terms of comoving variables \cite{Esposito_NuPhB2000}.

\subsection*{QED corrections}

QED corrections modify the mechanism presented before in several ways. In this paper, we only consider the changes to the thermodynamics of the plasma \cite{Heckler_PhRvD1994,Fornengo_PhRvD1997,Bennett2019}, since full corrections to the weak rates remain to be calculated and would correspond to a higher-order effect. In the following we use the notations of Ref.~\cite{Mangano2002}. Changes in the thermodynamics of the electromagnetic plasma induce a decrease in the total pressure,
\begin{equation}
P^\mathrm{int} = - \alpha \pi T_\gamma^4 \left( \frac23 K\left(\frac{m_e}{T_\gamma}\right) + 2 K\left(\frac{m_e}{T_\gamma}\right) ^2 \right) \, ,
\end{equation}
in agreement with Eq.~(48) of Ref.~\cite{Pitrou_2018PhysRept}. Note that we only kept the momentum-independent part of the electron mass shift derived in Ref.~\cite{Heckler_PhRvD1994}, as it is the dominant contribution. This result is also in agreement with the limit $m_e \to 0$ used in Ref.~\cite{Grohs_insights}. Using the classical thermodynamics relation $\rho = -P + T \dd P/\dd T$, we derive the energy density contribution corresponding to QED effects,
\begin{align}
\rho^\mathrm{int} &= \pi \alpha T_\gamma^4 \left( - 2 K - 6K^2 + \frac23 \frac{m_e}{T_\gamma} K' + 4 \frac{m_e}{T_\gamma} KK'\right) \\
&= \pi \alpha T_\gamma^4 \left( - \frac23 (K + J) + 2K(K-2J) \right) \, ,
\end{align}
where, for instance, $J$ stands for $J(m_e/T_\gamma)$.

Equation~\eqref{eq:z} is modified by these extra contributions \cite{Mangano2002}:
\begin{equation}
\frac{dz}{dx} =  \frac{\displaystyle \frac{x}{z}J(x/z) - \frac{1}{2 \pi^2 z^3} \frac{1}{xH} \int_{0}^{\infty}{\dd y \, y^3 \left[C_{\nu}\right]} + G_1(x/z)}{ \displaystyle \frac{x^2}{z^2}J(x/z) + Y(x/z) + \frac{2 \pi^2}{15} + G_2(x/z)} \, . \label{eq:zQED}
\end{equation}

The functions $G_1$ and $G_2$ are given in Eqs.~(18)-(19) of Ref.~\cite{Mangano2002} and Eqs.~(4.13)-(4.14) of Ref.~\cite{Bennett2019}. We found a simpler expression for $G_1$ which we reproduce here:
\begin{equation}
  G_1(\tau) = 2 \pi \alpha \left[ \frac{K'(\tau)}{3} + \frac{J'(\tau)}{6} + J'(\tau)K(\tau) + J(\tau)K'(\tau) \right]
 \end{equation}
This shows the advantage of not including a factor $1/\tau$, which is numerically challenging for high $T_\gamma$. It is actually equivalent to the expression in Refs.~\cite{Mangano2005,Bennett2019} through the relation
\begin{equation}
2 K (\tau) - \tau K'(\tau) = J(\tau) \, .
\end{equation}

\end{document}